\documentclass[%
superscriptaddress,
nofootinbib,
amsmath,amssymb,
 aps,
]{revtex4-1}
\usepackage{amsmath,amssymb,latexsym}
\usepackage{graphicx}
\usepackage{dcolumn}
\usepackage{bm,hyperref} 
\usepackage[table,dvipsnames]{xcolor}
\usepackage{physics}
\usepackage{natbib}
\usepackage[english]{babel}
\usepackage{mathtools}
\usepackage{soul}
\newcommand{\cspace}[2]{{}^{\star}\ensuremath{\mathbb{T}_{#1}\rm#2}}
\newcommand{\tspace}[2]{\ensuremath{\mathbb{T}_{#1}\rm#2}}
\newcommand{\tbundle}[1]{\ensuremath{\mathbb{T}\mathcal{#1}}}

\newcommand{\nn}{\nonumber}
\def\a{\alpha}
\def\b{\beta}

\def\d{\delta} 
\def\r{\rho}
\def\s{\sigma}

\newcommand{\p}{\partial}
\newcommand{\lieder}[2]{\mathcal{L}_{#1}{#2}}
\newcommand{\christoffel}[3]{\Gamma^{#1}_{#2#3}}

\newcommand{\inv}[1]{\frac{1}{#1}}
\def\epsLmnab{\epsilon_{\mu\nu\alpha\beta}}
\def\epsUmnab{\epsilon^{\mu\nu\alpha\beta}}

\newcommand{\diag}{\mathop{\rm diag}\nolimits}

\newcommand{\mcC}{{\mathcal{C}}}

\newcommand{\mcE}{{\mathcal{E}}}

\newcommand{\mcK}{{\mathcal{K}}}

\newcommand{\mcM}{{\mathcal{M}}}

\newcommand{\mcP}{{\mathcal{P}}}
\newcommand{\mcQ}{{\mathcal{Q}}}

\newcommand{\mcV}{{\mathcal{V}}}

\newcommand{\propt}{\mathfrak{s}}
\newcommand{\enp}{h}
\newcommand{\eps}{\varepsilon}
\newcommand{\vs}{v_s}
\newcommand{\hlift}{\mathcal{D}}
\newcommand{\chlift}{\Tilde{\mathcal{D}}}

\def\eqp{\mathrm{e}}

\def\eq{\mathrm{eq}}
\def\HL{HL}
\def\eqp{\mathrm{e}}

\def\eq{\mathrm{eq}}
\def\bulkr{R_\zeta}
\def\bulkc{C_\zeta}
\def\shr{R_\eta}
\def\shc{C_\eta}

\def\bulkrc{\mcV_\zeta}
\def\shrc{\mcV_\eta}
\newcommand{\adm}[1]{\mathcal{C}\left[#1\right]}
\newcommand{\half}{\frac{1}{2}}
\def\neqp{\mathrm{ne}}
\newcommand{\MeV}{\mathop{\rm MeV}}
\newcommand{\fm}{\mathop{\rm fm}}

\newcommand{\amm}[1]{#1}
\begin{document}

\title{Linear stability analysis in inhomogeneous equilibrium configurations}

\author{Masoud Shokri}
\affiliation{Institut f\"ur Theoretische Physik, 
	Johann Wolfgang Goethe--Universit\"at,
	Max-von-Laue-Str.\ 1, D--60438 Frankfurt am Main, Germany}

\author{Dirk H.\ Rischke}
\affiliation{Institut f\"ur Theoretische Physik, 
	Johann Wolfgang Goethe--Universit\"at,
	Max-von-Laue-Str.\ 1, D--60438 Frankfurt am Main, Germany}
\affiliation{Helmholtz Research Academy Hesse for FAIR, Campus Riedberg,\\
Max-von-Laue-Str.~12, D-60438 Frankfurt am Main, Germany}

\date{\today}

\begin{abstract}
	We propose a novel method to find local plane-wave solutions of the linearized equations of motion of relativistic hydrodynamics in inhomogeneous equilibrium configurations, i.e., when a fluid in equilibrium is rigidly moving with nonzero thermal vorticity. 
 Our method is based on extending the conserved currents to the tangent bundle, using a type of Wigner transformation.
 The Wigner-transformed conserved currents can then be Fourier-transformed into the cotangent bundle to obtain the dispersion relations for the space-time dependent eigenfrequencies. 
 We show that the connection between the stability of hydrodynamics and the evolution of plane waves is not as straightforward as in the homogeneous case, namely, it is restricted to the equilibrium-preserving directions in the cotangent bundle. 
 We apply this method to M\"uller-Israel-Stewart (MIS) theory and show that the interplay between the bulk viscous pressure and the shear-stress tensor with acceleration and rotation leads to novel modes, as well as modifications of the already known ones. 
 We conclude that, within the domain of applicability, i.e., when boundary effects are negligible and the vorticity is not too large, MIS theory is stable and causal, with the same stability and causality conditions as for homogeneous equilibrium configurations.
\end{abstract}

\maketitle

\section{Introduction}
\label{sec:intro}

Hydrodynamics is a theory that describes the long-wavelength behavior of fluids near local thermodynamical equilibrium \cite{rezzolla2013relativistic,denicol_rischke_2022}. 
Its equations of motion comprise the conservation of various currents, most importantly the energy-momentum tensor, as well as those of conserved charges in the system. 
More often than not, the form of the conserved currents is only rigorously known in equilibrium.
In such a state, the conserved currents are expressed in terms of hydrodynamic fields, such as the fluid four-velocity and temperature. 
For perfect fluids, knowing the equilibrium forms of the conserved currents is sufficient. 
However, real-world fluids experience dissipation. 
To describe them, we need to identify the relevant out-of-equilibrium contributions to the conserved currents. 
There are different ways to construct such terms. As is expected on physical grounds, some of these terms contain derivatives of the hydrodynamic fields, which gives rise to the so-called gradient expansion. 
One starts by assuming that, near equilibrium, the gradients of the fields are smaller than the fields themselves. Therefore, the additional terms in the conserved currents must comprise these gradients multiplied by parameters, the so-called transport coefficients, which define the responses of the fluids to these gradients.
These coefficients can be determined from an underlying theory that determines the microscopic dynamics of the system under consideration. 
At first order in derivatives, the gradient expansion yields Navier-Stokes theory \cite{LLfluid,Eckart:1940te}.

 It is legitimate to ask if the equilibrium state has maximum entropy in a hydrodynamic theory arising from the gradient expansion \cite{Gavassino:2020ubn}. 
This question is synonymous to the stability of hydrodynamics. 
It can be addressed by assuming an equilibrium state and asking if small perturbations remain small with increasing time. 
In this spirit, Hiscock and Lindblom (\HL{}) assumed plane-wave perturbations around a homogeneous equilibrium state and showed that such a state is indeed unstable in Navier-Stokes theory \cite{Hiscock:1985zz}. 
Also, it is well known that the equations of motion of Navier-Stokes theory are parabolic; therefore, they allow for the propagation of signals outside the causal light cone. 
In the plane-wave analysis of \HL{}, which we will refer to as linear stability analysis, this fact is exhibited in the existence of waves that, for short wavelengths, travel faster than light. 
They also found that some modes, which are damped in the frame of a comoving observer, are unstable in the frame of another observer, which is moving uniformly with a finite speed with respect to the fluid. 
This connection between stability and causality was also investigated in Refs.\ \cite{Denicol:2008ha,Pu:2009fj}, and was finally settled in Ref.\ \cite{Gavassino:2021owo}, where it was found that, in the linear regime, for a causal theory of hydrodynamics, damped modes remain damped in any inertial frame.

The instability of Navier-Stokes theory, which cannot be cured by including higher-order terms in the gradient expansion, was one of the main factors in the development of causal and stable theories of hydrodynamics. 
In particular, M\"uller-Israel-Stewart (MIS) \cite{muller,Israel:1976tn,Israel:1979wp} theory emerged as an answer to this problem. 
The stability of this theory, under certain conditions, was investigated both by linear stability analysis \cite{Hiscock:1987zz} and also by a method relying on Gibbs' stability criteria \cite{Hiscock:1983zz,Olson:1990rzl}, which was recently put into a systematic form in Ref.\ \cite{Gavassino:2021kjm}. 
Following this reference, we will call this method the information-current method. 
On the other hand, using kinetic theory, so-called Denicol-Niemi-Molnar-Rischke (DNMR) theory \cite{Denicol:2012cn} was developed, which in the linear regime is similar to MIS theory. 
Recently, it was discovered that a first-order stable and causal theory of hydrodynamics indeed exists if one does not use the standard matching conditions according to Landau \cite{LLfluid} or Eckart \cite{Eckart:1940te}. 
Such an improved gradient expansion gives rise to so-called Bemfica-Disconzi-Noronha-Kovtun (BDNK) theory of first-order hydrodynamics \cite{Bemfica:2017wps,Bemfica:2019knx,Bemfica:2020zjp,Kovtun:2019hdm,Hoult:2020eho}.

The linear stability analysis not only enables us to understand the stability of hydrodynamics but also reveals the nature of the waves arising from perturbing the equilibrium state. 
However, unlike the information-current method, it requires the existence of a homogeneous equilibrium configuration, i.e., a state where the hydrodynamic fields do not depend on space-time. 
On the other hand, the information-current method does not give us any information on the propagation of linear waves and is only applicable to theories for which the second law of thermodynamics holds exactly. 
This shortcoming is, in particular, relevant for BDNK theory, because its entropy current does not contain terms that ensure causality \cite{Kovtun:2019hdm}.
Furthermore, inhomogeneous equilibrium configurations, e.g., rigidly rotating fluids, always feature a length scale arising from the existence of a boundary, which is neglected in the information-current method.

It is known that the equilibrium configuration of an uncharged fluid is fully determined by a timelike Killing vector, which we refer to as $\beta$-vector (see, for example, Ref.\ \cite{Becattini:2016stj} and references therein). 
With the $\beta$-vector being fixed, the hydrodynamic variables, such as the four-velocity and temperature, are unambiguously determined. 
In a sense, one might say that geometry dictates the possible equilibrium configurations. 
Even in flat space-time, it is possible to have inhomogeneous equilibrium configurations. 
This, for example, includes the case of rigidly rotating fluids in equilibrium. 
Such equilibrium conditions have attracted attention in recent years in the context of heavy-ion physics, mainly due to the increasing interest in understanding the process of conversion of the orbital angular momentum in noncentral collisions into the polarization of observed particles \cite{Becattini:2020sww}.

Naturally, one may inquire if linear waves can also be found in an inhomogeneous equilibrium configuration. 
If yes, what can we then learn from them about the stability of the theory? 
In the current work, we will answer these questions. 
This paper is organized as follows: In Sec.\ \ref{sec:pre}, we review possible equilibrium configurations and the linear stability analysis. 
Then, in Sec.\ \ref{sec:tangent-bundle}, we develop the tools necessary to solve the linearized hydrodynamic equations of motion in inhomogeneous equilibrium configurations, at hand of the example of a simple wave equation. 
Namely, we extend the wave equation to the tangent bundle, using a kind of Wigner transformation.
 The solution of this extended wave equation is then Fourier-transformed into the cotangent bundle to find the dispersion relations for the space-time dependent eigenfrequencies. 
 We show that the connection between the stability of the solutions and the imaginary parts of the eigenfrequencies is restricted to the equilibrium-preserving directions in the cotangent bundle. 
 In Sec.\ \ref{sec:app_hyd} we apply these ideas to hydrodynamics in general.
Subsequently, in Sec.\ \ref{sec:mis}, we determine the modes of MIS hydrodynamics in inhomogeneous equilibrium configurations and investigate the interplay between dissipative fluxes, acceleration, and rotation. 
Section \ref{sec:conclusion} concludes this paper with a summary of our results and an outlook. 
Details of our calculations are delegated to several appendices.

%%%%%%%%%%%%%%%%%%%%%
\paragraph*{Notations and conventions}
We use natural units $\hbar=c=k=1$.
Euclidean three-vectors are denoted with boldface letters, such as $\vb{y}$, in contrast to four-vectors, like $y$. 
The index-free notation is often used for four-vectors, for example, $u=u^\mu\p_\mu$. 
We use the dot notation for scalar products, both between four and three-vectors, i.e., $a\cdot b = a^\mu b_\mu$ and $\vb{a}\cdot\vb{b}$.
The covariant and Lie derivatives are denoted by $\nabla$ and $\lieder{}{}$, respectively. 
We denote the horizontally lifted covariant derivative with $\hlift{}$ in the tangent bundle and $\chlift{}$ in the cotangent one. 
The metric signature is mostly minus, i.e., $\eta_{\mu \nu} = \diag(1,-1,-1,-1)$. 
Our convention for the totally antisymmetric tensor $\epsUmnab$ is such that in Minkowskian coordinates $\eps^{0123}=-\eps_{0123}=1$. 
We use the standard symmetrization and antisymmetrization notations, $A_{(\mu\nu)} \equiv \tfrac{1}{2}\left(A_{\mu\nu} +A_{\nu\mu}\right)$ and
$A_{[\mu\nu]} \equiv \tfrac{1}{2}\left(A_{\mu\nu} -A_{\nu\mu}\right)$, respectively.
The covariant projector $\Delta^{\mu\nu} \equiv g^{\mu\nu}-u^\mu u^\nu$, with $u^\mu$ being the fluid four-velocity, projects every vector $A^\mu$ onto the three-space orthogonal to $u^\mu$, i.e., $A^{\langle\mu\rangle} \equiv \Delta^{\mu\nu}A_\nu$. 
The symmetric, traceless projector of rank four is $\Delta^{\mu\nu}_{\alpha\beta} \equiv \tfrac{1}{2}\left(\Delta^{\mu}_{\alpha}\Delta^{\nu}_{\beta}+\Delta^{\mu}_{\beta}\Delta^{\nu}_{\alpha}\right)-\tfrac{1}{3}\Delta^{\mu\nu}\Delta_{\alpha\beta}$, the application of which onto a rank-2 tensor $A^{\mu\nu}$ is denoted by $A^{\langle\mu\nu\rangle} \equiv \Delta^{\mu\nu}_{\alpha\beta}A^{\alpha\beta}$.
The convention that we use for the Riemann tensor is $R^\sigma_{\r\mu\nu} = 2\left(\partial_{[\mu}\christoffel{\sigma}{\nu]}{\rho}
+\christoffel{\sigma}{[\mu}{\beta}\christoffel{\beta}{\nu]}{\r}\right)$.

\section{\label{sec:pre}Preliminaries}

In this section, we briefly review the concepts required for the remainder of this work. 
Let us consider a fluid described by a set of conserved currents $\{Q^{\mu\nu\cdots}_1, Q^{\mu\nu\cdots}_2, \ldots \}$. 
We refer to the conservation equations satisfied by these currents, i.e., $\nabla_\mu Q^{\mu\nu\cdots}_i = 0 $, with $i= 1,2, \ldots$, as equations of motion (EOM). 
Although the system in consideration may possess multiple conserved currents, for the following we assume a neutral simple fluid that only has the energy-momentum tensor $T^{\mu\nu}$ as conserved current. 

In global equilibrium, there exists a timelike Killing vector $\beta^\mu$ (see, e.g., Ref.~\cite{Becattini:2016stj} for a review), i.e.,
\begin{equation}\label{eq:killing}
	\lieder{\beta}{g_{\mu\nu}} = \nabla_\mu \beta_\nu + \nabla_\nu \beta_\mu = 0\,,\qq{and}\beta \cdot \beta > 0\,.
\end{equation}
from which the fluid's four-velocity and temperature can be computed as
\begin{equation}\label{eq:def-tu}
	u^\mu = \frac{\beta^\mu}{\sqrt{\beta \cdot \beta}}\,,\qquad
	T = \frac{1}{\sqrt{\beta \cdot \beta}}\,.
\end{equation}
A fluid in global equilibrium does not necessarily move with a uniform velocity, in fact, it can be subject to global rotation and/or acceleration. Such nontrivial kinematics can be encoded in an antisymmetric rank-2 tensor, which is referred to as the thermal vorticity,
\begin{equation}\label{eq:tvort}
	\varpi_{\mu\nu} \equiv -\nabla_{[\mu}\b_{\nu]}\,.
\end{equation}
As an antisymmetric rank-2 tensor field, $\varpi_{\mu \nu}$ can be decomposed as
\begin{equation} \label{eq:vorticity}
\varpi_{\mu \nu} = \frac{2}{T} a_{[\mu} u_{\nu]}
+ \frac{1}{T} \epsilon_{\mu \nu \alpha \beta}
\omega^\alpha u^\beta\,,
\end{equation}
where $a_\mu \equiv  T \varpi_{\mu\nu}u^\nu$ is the \textit{electric} part of the thermal vorticity and $\omega^\mu \equiv -\inv{2}T \epsUmnab u_\nu \varpi_{\a\b}$ is the \textit{magnetic} part.

Using Eqs.\ \eqref{eq:killing} and \eqref{eq:def-tu} one finds that both temperature and four-velocity commute with $\beta^\mu$, i.e., their Lie derivatives with respect to $\beta$ vanish.
This is in fact a general result: any physical quantity described by a tensor $X^{\mu\nu\cdots}$ of arbitrary rank commutes with $\beta^\mu$ in global equilibrium, namely, \cite{Becattini:2016stj}
\begin{equation}\label{eq:lie-physics}
	\lieder{\beta}{X^{\mu\nu\cdots}} = 0\,.
\end{equation}
Using $\lieder{\beta}{T}=0$, we find that $a_\mu \equiv u \cdot \nabla \, u_\mu$ is the \textit{four-acceleration} of the fluid, while $\omega^\mu$ is usually referred to as the \textit{kinematic vorticity four-vector}. 
Note that $\omega^\mu = \frac{1}{2}  \epsilon^{\mu \nu \alpha \beta} u_\nu \Omega_{\alpha \beta}$,
where $\Omega_{\alpha \beta} \equiv \half \left( \nabla_{\langle\alpha\rangle} u_{\beta} - \nabla_{\langle \beta \rangle} u_\a \right)$ is the rank-2 \textit{fluid vorticity tensor}. 
Moreover, the acceleration and the gradient of temperature are related through
\begin{equation}\label{eq:a-grad-T}
	T a_\mu = \nabla_\mu T\,.
\end{equation}

The hydrodynamic fields that arise from Eqs.\ \eqref{eq:def-tu} and \eqref{eq:lie-physics}, with $\beta^\mu$ being a Killing vector, satisfy the perfect-fluid EOM, i.e., $\nabla_\mu T^{\mu \nu}_\eq =0$.
Furthermore, dissipative currents must be constructed such that they vanish in global equilibrium regardless of the relevant transport coefficients. 
Thus, the conserved currents reduce to their perfect-fluid counterparts, and the EOM are guaranteed to be satisfied in equilibrium.\footnote{In the case of nonvanishing curvature, a derivative expansion of the conserved currents also features terms which contain derivatives of the metric. These curvature-induced terms do not vanish in equilibrium and are thus not of dissipative nature. Nevertheless, an equilibrium configuration defined via a timelike Killing vector remains a solution to the EOM  \cite{Kovtun:2022vas}.} 

\subsection{Homogeneous and inhomogeneous equilibrium configurations}
\label{sec:eq-cats}

At this stage, let us review some features of possible equilibrium configurations, and categorize them.
In Minkowski space-time, the vector
\begin{equation}\label{eq:hydrostatic}
	\beta = \frac{1}{T_0}\pdv{t}\,,   
\end{equation}
with a positive constant $T_0$, is a timelike Killing vector. 
Using $\beta \equiv \beta \cdot \partial$, this $\beta$-vector corresponds to a fluid at rest with a global constant temperature, 
\begin{equation}\label{eq:hydrostatics}
	u^\mu = \left(1,\vb{0}\right)\,,\qquad T(t,\vb{x}) = T_0\,,
\end{equation}
i.e., the fluid is in \textit{hydrostatic equilibrium}.

Adding a Killing vector to a Killing vector yields by definition another Killing vector.
If the sum is timelike, it can be regarded as the $\beta$-vector for another possible equilibrium configuration. 
In general, Minkowski space-time possesses ten independent Killing vectors, corresponding to the generators of the Poincar\'{e} algebra, i.e., in addition to $\pdv{t}$, the generators of the three spatial translations, $\pdv{x^i}$, the three spatial rotations, $\epsilon^{ijk} x^j \pdv{x^k}$, and the three Lorentz boosts, $ x^i \pdv{t}+t \pdv{x^i}$, where $i, j, k = 1,2,3$.  
Therefore, adding $T_0^{-1} v^i \pdv{x^i}$ to Eq.\ (\ref{eq:hydrostatic}) results in a timelike Killing vector if the modulus of the coefficient $v^i$ fulfills $ |v^i| < 1$,
\begin{equation}\label{eq:moving-hydrostatic}
	\beta = \inv{T_0}\left(\pdv{t}+v^i\pdv{x^i}\right)\,.
\end{equation}
Summing over $i$, we obtain a Killing vector if $\sum_{i=1}^3 (v^i)^2 <1$.  
With the definitions \eqref{eq:def-tu}, one then obtains
\begin{equation}\label{eq:hs-vt}
	u^\mu = \gamma\left(1,\vb{v}\right)\,,\qquad T = \gamma T_0\,,
\end{equation}
i.e., $v^i$ are the components of the three-velocity $\vb{v}$, with $\gamma \equiv 1/\sqrt{1-\vb{v}^2}$ being the Lorentz factor. 
Configurations \eqref{eq:hydrostatic} and \eqref{eq:moving-hydrostatic} are related through a global, i.e., space-time-independent, boost. In all these cases, physical quantities are constant in space-time. Thus, we refer to such configurations as \textit{homogeneous equilibrium configurations}.

However, even in flat space-time, \textit{inhomogeneous equilibrium configurations} are possible, for which the hydrodynamic quantities are not constant in space and time.
These are found by adding the generators of boosts and rotations to the hydrostatic $\beta$-vector \eqref{eq:hydrostatic}. 
For example, by adding the generator of a boost along the $z$-direction, multiplied with a coefficient $a_0/T_0$, where $a_0$ is a positive constant of dimension energy, the $\beta$-vector assumes the form~\cite{Palermo:2021hlf}
\begin{equation}\label{eq:acc-conf}
	\beta = \inv{T_0}\left[\pdv{t}+a_0\left(z\pdv{t}+t\pdv{z}\right)\right]\,.
\end{equation}
For $\beta$ to be timelike, it is required that
\begin{equation}
	\abs{1+a_0 z} > \abs{a_0 t}\,.
\end{equation}
It is simpler to express this configuration in so-called Rindler coordinates $(\tau,x,y,\xi)$, which are related to Minkowski coordinates through
\begin{align}\label{eq:rindler}
	\tau & = \frac{1}{2a_0}\log\left[\frac{1+a_0\left(z+t\right)}{1+a_0\left(z-t\right)}\right]\,, \qquad \xi  = \inv{2a_0}\log\left[\left(1+a_0z\right)^2-a_0^2t^2\right]\,.
\end{align}
The line element in the above coordinates reads
\begin{equation}
	\dd{s}^2=e^{2a_0\xi}\left(\dd{\tau}^2-\dd{\xi}^2\right)-\dd{x}^2-\dd{y}^2\,.\label{eq:ds2}
\end{equation}
Using the coordinate transformations \eqref{eq:rindler}, the $\beta$-vector has the simple form $\inv{T_0}\pdv{\tau}$, and the four-velocity (in Rindler coordinates) and temperature are obtained from Eq.\ \eqref{eq:def-tu} as
\begin{equation} \label{eq:umu_acc}
	u^\mu = e^{-a_0\xi}\, (1, \vb{0})\,,\qquad
	T = e^{-a_0\xi} \,T_0\,.
\end{equation}
We note that in Minkowski coordinates the four-velocity reads 
\begin{equation}
u^\mu = \gamma(t,z) \left(1,\vb{v}(t,z)\right)\,,
\qq{with} 
	\gamma = \cosh(a_0\tau)\,,\quad \vb{v} = \tanh(a_0\tau)\,\hat{\vb{z}}\,.
\end{equation}
This configuration has a nonzero acceleration,
which reads in Rindler coordinates
\begin{equation}
	a^\mu = a_0 e^{-2a_0\xi} \, (0,0,0,1)\,.
\end{equation}
The acceleration introduces a specific spacelike direction in equilibrium, which may be identified with the unit vector (in Rindler coordinates)
\begin{equation} \label{eq:lmu_acc}
	\ell^\mu = \frac{1}{a} a^\mu = e^{-a_0\xi} \, (0,0,0,1)\,,
\end{equation}
where $a = \sqrt{-a \cdot a}$.
We note that due to Eq.\ \eqref{eq:a-grad-T} the hypersurfaces perpendicular to $\ell^\mu$ are hypersurfaces of constant temperature. 
One convinces oneself that, for the configuration \eqref{eq:acc-conf}, the thermal vorticity does not have a magnetic, i.e., rotational, part. Therefore, we refer to this configuration as an \textit{accelerating configuration}. 
More general accelerating configurations can be found by adding the boost generators in $x$- and $y$-directions, multiplied with appropriate constant factors, to Eq.\ \eqref{eq:acc-conf}, respecting the restriction that the resulting $\beta$-vector is timelike.

Another inhomogeneous equilibrium configuration can be obtained by adding a generator of a rotation, multiplied with a coefficient $\Omega_0 /T_0$, where $\Omega_0$ is a positive constant with dimension energy, to the hydrostatic $\beta$-vector \eqref{eq:hydrostatic}. 
For instance, for a rotation around the $z$-axis, we then obtain~\cite{Chernodub:2016kxh,Palermo:2021hlf}
\begin{equation}\label{eq:rigid-rotation}
	\beta = \inv{T_0}\left[\pdv{t}+\Omega_0\left(x\pdv{y}-y\pdv{x}\right)\right]\,.
\end{equation}
This $\beta$-vector is timelike if
\begin{equation}
	\Omega_0^2 \left(x^2+y^2\right) < 1\,.
\end{equation}
This equilibrium configuration corresponds to a rigid rotation around the $z$-axis, wherefore
we call it a \textit{rotating  configuration}. It can be expressed in a simpler way in cylindrical coordinates $(t,\rho,\varphi,z)$, where $\rho = \sqrt{x^2+y^2},\,\varphi = \arctan(y/x)$, where the line element is
\begin{equation}
	\dd{s}^2= \dd{t}^2-\dd{\rho}^2-\rho^2\dd{\varphi}^2-\dd{z}^2\,.
\end{equation}
Using Eq.\ \eqref{eq:def-tu},
the four-velocity (in cylindrical coordinates)
and the temperature are obtained as
\begin{equation} \label{eq:umu_rot}
	u^\mu = \gamma(\rho)\, \left(1, 0, \Omega_0,0 \right)\,,\qquad 
	T = \gamma(\rho)\, T_0\,, \quad
\qq{with} 
	\gamma (\rho) = \frac{1}{\sqrt{1-\rho^2\Omega_0^2}}\,.    
\end{equation}
In this case, the thermal vorticity has both electric and magnetic parts, encoded in the acceleration and kinematic vorticity, which in cylindrical coordinates read
\begin{equation} \label{eq:amu_rot}
	a^\mu = -\gamma^2(\rho)\, \rho \Omega_0^2 \, (0, 1, 0,0)\,,\qquad 	
 \omega^\mu = \gamma^2(\rho)\, \Omega_0 \,  (0,0,0,1)\,,
\end{equation}
respectively. 
Although these vectors are orthogonal in this case, this is not a general result for all rotating configurations, see Appendix\ \ref{app:Example}.

As will become clear later, in the rotating case it is advantageous to define a tetrad of orthogonal four-vectors.
Obviously, $u^\mu$ is orthogonal to both $a^\mu$ and $\omega^\mu$, but the latter two are not necessarily orthogonal to each other.
Therefore, we decompose $\omega^\mu$ into directions parallel and orthogonal to the normalized acceleration $\ell^\mu$, 
\begin{equation} \label{eq:decomp_vort}
   \omega^\mu = \omega_\ell \ell^\mu + \omega_\perp \psi^\mu\,,
\end{equation} 
with $\omega_\ell \equiv - \ell \cdot \omega$, $\omega_\perp = \sqrt{-\omega \cdot \omega -\omega_\ell^2}$, and $\psi^\mu \equiv (\omega^\mu- \omega_\ell \ell^\mu )/\omega_\perp$. 
Note that $\psi^\mu$ is only well defined when $\omega_\perp \neq 0$, which is always fulfilled for the rotating configuration.
Then we define   
\begin{equation}\label{eq:def-zeta}
	\zeta_\mu \equiv \epsLmnab u^\nu \ell^\a \, \psi^\b\,.
\end{equation}
For the rotation around the $z$-axis $\zeta^\mu$ reads in cylindrical coordinates
\begin{equation} \label{eq:zetamu_rot}
	\zeta^\mu = - \frac{\gamma(\rho)}{\rho}\, \left(\rho^2\Omega_0, 0, 1,0\right)\,.
\end{equation}
The set of vectors $(u,\ell,\psi,\zeta)$
then forms a tetrad of orthonormal four-vectors.

We can combine the rotating and accelerating cases with each other or with the homogeneous case to find more complicated global-equilibrium configurations. 
Also, rotations may occur around different axes.
One may also assume a curved background. An example is given in Appendix\ \ref{app:Example}.

\subsection{Linear stability of homogeneous equilibrium configurations}\label{sec:lin-stab-rev}

As mentioned in the Introduction, our goal is to generalize the linear stability analysis of hydrodynamic theories to inhomogeneous equilibrium configurations. 
It is therefore useful to first remind ourselves of the standard linear stability analysis in homogeneous equilibrium configurations \cite{Hiscock:1985zz}. 
For a homogeneous equilibrium configuration, the fluid moves with a four-velocity corresponding to the $\beta$-vector  \eqref{eq:moving-hydrostatic} in an observer's frame. 
The four-velocity of this observer defines a timelike vector $n^\mu$.
In the observer's rest frame, $n^\mu  = (1,\vb{0})$ is the normal vector on a spacelike hypersurface $\Sigma(t)$ with volume element $\dd[3]{x}$, where $t$ is the time coordinate in the observer's frame. 
The energy-momentum tensor $T^{\mu\nu}$ is then perturbed with respect to its equilibrium value.
The perturbation $\delta T^{\mu \nu}$ is assumed to be small, such that the EOM can be linearized to first order in $\delta T^{\mu \nu}$.
		
In the following, we denote the components of $\delta T^{\mu \nu}$ as $\d X^A(t,\vb{x})$, with $A$ being the component index. 
Inserting $\d X^A(t,\vb{x})$ into the linearized EOM and solving the latter in Fourier space gives rise to a set of homogeneous linear equations for the Fourier components $\d X^A(\omega,\vb{k})$,
\begin{equation}
	\label{eq:linear-set}
	M^{A B}(\omega,\vb{k}) \d X^{B}(\omega,\vb{k})=0\,.
\end{equation}
This system has nontrivial solutions if the determinant of $M(\omega,\vb{k})$ vanishes. The (in general complex) roots of the characteristic equation $\det M(\omega,\vb{k})=0$ give the dispersion relations of the normal modes of the system
\begin{equation}\label{eq:dispresion}
	\omega_a = \omega_a(\vb{k})\,,
\end{equation}
where $a$ labels the various modes. 
A mode becomes unstable if (in our convention for the Fourier transformation) $\mathrm{Im} \,\omega_a(\vb{k})>0$ in some domain $D_{\vb{k}}$ of the space of three-momenta $\vb{k}$. 
One can show that if at least one mode is unstable, the $L^2$ norm 
\begin{equation}\label{eq:flat-norm}
	\norm{\d X^A(t)}^2 = \int_{\Sigma(t)} \dd[3]{x}\left|\int_{\vb{k}}\sum_a \d X^A_a(\vb{k})e^{-i\omega_a(\vb{k}) t + i \vb{k} \cdot \vb{x}}\right|^2\,,
\end{equation}
on spatial surfaces $\Sigma(t)$ diverges as $t \to \infty$. 
Here
\[
\int_{\vb{k}} \equiv \int\frac{\dd[3]{k}}{(2\pi)^3}\,.
\]
Vice versa, the equilibrium configuration is linearly stable if $\mathrm{Im}\, \omega_a(\vb{k})\leq 0$ for all modes and all values of $\vb{k}$. 

\subsection{Wave equation as an example}\label{sec:wave-hom}

In this subsection, we want to elucidate the concepts of the previous subsection at hand of a simple example: a relativistic wave equation of the form \cite{Friedlander:2010eqa}
\begin{equation}\label{eq:wave}
	\left( \Box - f \beta \cdot \nabla + m^2\right) \phi(x)= 0\,,   
\end{equation}
where $\Box \equiv \nabla \cdot \nabla$ is the d'Alembert operator, $f$ and $m$ are some coefficients, and $\beta^\mu$ is a timelike Killing vector. 
A homogeneous equilibrium configuration corresponds to the condition that $f$ and $m$ are constants and space-time is flat, while in an inhomogeneous equilibrium configuration, $f$ and $m$ are functions of space-time and/or space-time has a nontrivial curvature. 

In Minkowskian space-time, after Fourier transformation of Eq.\ \eqref{eq:wave}, we find the following \emph{characteristic equation} in some observer's frame
\begin{equation}\label{eq:wave-dispersion}
	\omega^2 - if\beta^0 \omega + if\boldsymbol{\beta}\cdot\vb{k} - \vb{k}^2 - m^2 = 0\,,
\end{equation}
where we have used that $\beta^\mu=\left(\beta^0,\boldsymbol{\beta}\right)$. 
The two roots of this equation $\omega_\pm(\vb{k})$ determine the \emph{dispersion relations}, and the solution of the wave equation (\ref{eq:wave}) is
\begin{equation}
\phi(t,\vb{x}) = \int_{\vb{k}}
\left[ \phi_+(\vb{k})
\, e^{-i \omega_+(\vb{k}) t +
i \vb{k} \cdot \vb{x}}
+ \phi_-(\vb{k})
\, e^{-i \omega_-(\vb{k}) t +
i \vb{k} \cdot \vb{x}} \right]\,.
\end{equation}
If $f > 0$, then the imaginary part of one of the roots, say $\omega_+(\vb{k})$, is positive in a subdomain $D_{\vb{k}}$ of the space of three-momenta $\vb{k}$. 
For example, if $m=0$, there are two roots for $\vb{k} = 0$, $\omega_+(\vb{0}) = if\beta_0, \omega_-(\vb{0}) = 0$. 
Following Ref.~\cite{Hiscock:1985zz}, we can then show that the $L^2$ norm satisfies
\begin{equation} \label{eq:norm_unstable}
	\norm{\phi(t)}^2 \geq e^{2\Lambda t} \int_{\vb{k} \in D_{\vb{k}}}\left|\phi_+(\vb{k})+\phi_-(\vb{k})e^{-\varsigma t + i \Re\Delta \omega(\vb{k}) t}\right|^2\,,
\end{equation}
where $\Lambda$ is the minimum value of $\Im\omega_+(\vb{k})$ on $D_{\vb{k}}$, $\Delta\omega (\vb{k})= \omega_+(\vb{k}) -\omega_-(\vb{k})$, and $\varsigma$ is the maximum value of $\Im\Delta\omega(\vb{k})$ on $D_{\vb{k}}$. 
This inequality shows that the norm is growing unboundedly with time. 

Now, let us assume an inhomogeneous equilibrium configuration, for which $m$ and $f$ are only constant on integral curves of the Killing vector $\beta^\mu$, namely where
\begin{equation} \label{eq:integral_curves}
	\lieder{\beta}{m} = \lieder{\beta}{f} =0 \,.
\end{equation}
Now we repeat the same procedure as above, i.e., we perform a Fourier transformation of Eq.\ \eqref{eq:wave}. 
However, now $m$ and $f$ are in general not constant on $\Sigma(t)$ (apart from the lower-dimensional manifold defined by Eq.\ \eqref{eq:integral_curves}), and the characteristic equation \eqref{eq:wave-dispersion}, and thus the dispersion relations, will also be coordinate-dependent. This
is inconsistent with replacing derivatives $\partial_\mu$ by $-ik_\mu$, even in flat space-time, and the wave equation \eqref{eq:wave} cannot be solved by Fourier transformation.
In the following section, we propose an approach to handle this problem at least in flat space-time.

\section{Extension to the tangent bundle}\label{sec:tangent-bundle}

In this section, we propose a procedure that can be used for the linear stability analysis in an inhomogeneous equilibrium configuration in flat space-time. 
Inspired by quantum transport theory in curved space-time \cite{fonarev}, we extend the perturbations to the tangent bundle using a so-called Wigner transform. 
We then study the wave equation \eqref{eq:wave} in tangent space. 
Analyzing the stability of its solutions requires a restriction of the norm to the equilibrium-preserving directions in tangent space. 
We first define the latter and then apply this concept to the definition of the norm.

\subsection{The Wigner transform and its properties} 
Let $F^{\mu_1 \mu_2 \cdots}_{\nu_1 \nu_2 \cdots}$ be a tensor field of arbitrary rank defined in some arbitrary Lorentzian space-time manifold $\mcM$, and $y$ a tangent vector at a point $\mcP \in \mcM$ with coordinates $x$. 
Then, following Ref.\ \cite{fonarev}, we call the following construction the Wigner transform of $F^{\mu_1 \mu_2 \cdots}_{\nu_1 \nu_2 \cdots}$,
\begin{equation}  \label{eq:def_Wigner}
	F^{\mu_1 \mu_2 \cdots}_{\nu_1 \nu_2 \cdots}(x,y) \equiv e^{y \cdot \hlift} F^{\mu_1 \mu_2 \cdots}_{\nu_1 \nu_2 \cdots}(x)\,,
\end{equation}
where $\hlift_\a \equiv \nabla_\a - \christoffel{\s}{\a}{\r}y^\r\p_\s^y$ is the horizontal lift in the tangent bundle $\tbundle{M}$. 
Note that the explicit form of the covariant derivative $\nabla$ in $\hlift$ depends on the tensor rank of $F$, but the second part of $\hlift_\a$ does not. 
To recover the \textit{base} tensor $F^{\mu_1 \mu_2 \cdots}_{\nu_1 \nu_2 \cdots}(x)$ from its Wigner transform, one only needs to evaluate the latter at $y=0$, i.e., 
\begin{equation}\label{eq:wigner-delta}
	F^{\mu_1 \mu_2 \cdots}_{\nu_1 \nu_2 \cdots}(x) = \int_{\tspace{x}{M}}\dd[4]{y} \d^{4}(y)  F^{\mu_1 \mu_2 \cdots}_{\nu_1 \nu_2 \cdots}(x,y)\,,
\end{equation}
where $\tspace{x}{M}$ denotes the tangent space at point $\mcP$. 
Since the tangent space is Minkowskian, we may Fourier-transform the Dirac delta function to obtain
\begin{equation} 
	F^{\mu_1 \mu_2 \cdots}_{\nu_1 \nu_2 \cdots}(x) = \int_{\tspace{x}{M}}\dd[4]{y} \int_{\cspace{x}{M}} \frac{\dd[4]{k}}{(2\pi)^4} e^{i k \cdot y}  F^{\mu_1 \mu_2 \cdots}_{\nu_1 \nu_2 \cdots}(x,y)\,,
\end{equation}
where $k_\mu$ is an element of the cotangent space $\cspace{x}{M}$, and hence $k\cdot y$ is a scalar under coordinate transformations. 
The above relation implies the following definition of the Fourier transform of $F^{\mu_1 \mu_2 \cdots}_{\nu_1 \nu_2 \cdots}(x,y)$, 
\begin{equation} \label{eq:F}
	F^{\mu_1 \mu_2 \cdots}_{\nu_1 \nu_2 \cdots}(x,k) = \int_{\tspace{x}{M}}\dd[4]{y} \sqrt{-g} \, e^{i k \cdot y}  F^{\mu_1 \mu_2 \cdots}_{\nu_1 \nu_2 \cdots}(x,y)\,,
\end{equation}
and its inverse,
\begin{equation} \label{eq:F_inverse}
	F^{\mu_1 \mu_2 \cdots}_{\nu_1 \nu_2 \cdots}(x,y) = \int_k e^{-i k \cdot y}  F^{\mu_1 \mu_2 \cdots}_{\nu_1 \nu_2 \cdots}(x,k)\,,
\end{equation}
where
\begin{equation} \label{eq:measure}
	\int_k \equiv 
	\int_{\cspace{x}{M}}\frac{\dd[4]{k}}{\sqrt{-g}(2\pi)^4}\,,
\end{equation}
and where the square root of the metric determinant $\sqrt{-g}$ 
in Eqs.\ \eqref{eq:F} and \eqref{eq:measure} is required to render the integration measures scalars under coordinate transformations.
Inserting Eq.\ \eqref{eq:F_inverse} into Eq.\ \eqref{eq:wigner-delta} implies that
\begin{equation} \label{eq:useful_id}
    F^{\mu_1 \mu_2 \cdots}_{\nu_1 \nu_2 \cdots}(x) = \int_k F^{\mu_1 \mu_2 \cdots}_{\nu_1 \nu_2 \cdots}(x,k)\,.
\end{equation}

The covariant derivative of the base tensor field is related to the $y$-derivative of the Wigner transform in the following way,
\begin{eqnarray}\label{eq:dwigner-delta}
	\nabla_\mu  F^{\mu_1 \mu_2 \cdots}_{\nu_1 \nu_2 \cdots}(x) &=& \int_{\tspace{x}{M}}\dd[4]{y} \d^{4}(y) \p^y_\mu F^{\mu_1 \mu_2 \cdots}_{\nu_1 \nu_2 \cdots}(x,y)
	\\\nn
	&=& -\int_{\tspace{x}{M}}\dd[4]{y}   F^{\mu_1 \mu_2 \cdots}_{\nu_1 \nu_2 \cdots}(x,y) \p^y_\mu\d^{4}(y)\,.
\end{eqnarray}
The first line can be proven using the definition \eqref{eq:def_Wigner} of the Wigner transform under the integral on the right-hand side and employing the fact that, on account of the delta-function, only the term linear in $y$ of the Taylor expansion of $e^{y \cdot \hlift}$ survives.
In the first line of Eq.\ \eqref{eq:dwigner-delta}, one can replace the $y$-derivative with the horizontal lift in $\tbundle{M}$ using an important identity, which is proven in Appendix\ \ref{app:h-lift}, 
\begin{eqnarray}\label{eq:yd-to-hl}
	\p^y_\mu F^{\mu_1 \mu_2 \cdots}_{\nu_1 \nu_2 \cdots}(x,y) &=& \hlift_\mu F^{\mu_1 \mu_2 \cdots}_{\nu_1 \nu_2 \cdots}(x,y)
	- y^\nu \sum_{l=0}^{\infty}\frac{\mcC\left[y\cdot\hlift\right]^l}{\left(l+2\right)!}G_{\mu\nu}(x,y) F^{\mu_1 \mu_2 \cdots}_{\nu_1 \nu_2 \cdots}(x,y)\,.
\end{eqnarray}
Here, $\mcC[A]B \equiv \left[A,B\right]$ is the adjoint map and 
\begin{equation}\label{eq:def_G}
	G_{\mu\nu}(x,y) \equiv - R^\sigma_{\rho\mu\nu}y^\rho \p^y_\sigma\,.
\end{equation}
On the other hand, one can use the Fourier representation of the delta-function in the second line of Eq.\ \eqref{eq:dwigner-delta} and Eq.\ \eqref{eq:F} to obtain 
\begin{equation}\label{eq:dwigner-delta-cb}
	\nabla_\mu  F^{\mu_1 \mu_2 \cdots}_{\nu_1 \nu_2 \cdots}(x)= -i\int_k k_\mu F^{\mu_1 \mu_2 \cdots}_{\nu_1 \nu_2 \cdots}(x,k)\,.
\end{equation}

We also need to examine the horizontal lift in the cotangent bundle, i.e., $\chlift_\mu  \equiv \nabla_\mu  + \christoffel{\r}{\mu}{\sigma}k_\r \p_k^\sigma $. 
To this end, we start with
\begin{equation}\label{eq:hlift-to-chlift}
	\hlift_\mu F^{\mu_1 \mu_2 \cdots}_{\nu_1 \nu_2 \cdots}(x,y) = \int_k e^{-ik\cdot y}\chlift_\mu F^{\mu_1 \mu_2 \cdots}_{\nu_1 \nu_2 \cdots}(x,k) \,,
\end{equation}
which can be verified by noticing that the right-hand side subtracted from the left-hand side is a tensor that vanishes in the locally flat neighborhood of $\mcP$ \cite{fonarev}. 
Fourier-transforming this equation and employing Eq.\ \eqref{eq:yd-to-hl}, an integration by parts, and then Eq.\ \eqref{eq:F}, we obtain
\begin{equation}\label{eq:chlift-curved}
	\chlift_\mu F^{\mu_1 \mu_2 \cdots}_{\nu_1 \nu_2 \cdots}(x,k) = -i k_\mu F^{\mu_1 \mu_2 \cdots}_{\nu_1 \nu_2 \cdots}(x,k) + \text{curvature terms}\,.
\end{equation}
The curvature terms can be derived using
\begin{equation} \label{eq:G-F}
	G_{\mu\nu} (x,y) F^{\mu_1 \mu_2 \cdots}_{\nu_1 \nu_2 \cdots}(x,y) = \int_k e^{-ik\cdot y}\tilde{G}_{\mu\nu} (x,k) F^{\mu_1 \mu_2 \cdots}_{\nu_1 \nu_2 \cdots}(x,k) \,,
\end{equation}
with
$ \tilde{G}_{\mu\nu}(x,k) \equiv R^\sigma_{\rho\mu\nu}k_\sigma \partial_k^\rho$. Equation \eqref{eq:G-F} can be proved using Eqs.\ \eqref{eq:F} and \eqref{eq:def_G} and replacing $\partial_k^\rho \rightarrow i y^\rho$ and $k_\sigma \rightarrow i \partial_\sigma^y$.

\subsection{The wave equation in tangent space}\label{sec:ih-wave}
Let us now consider the wave equation \eqref{eq:wave} at some point $\mcP$ with coordinates $x$.
We then use Eqs.\ \eqref{eq:wigner-delta} and \eqref{eq:dwigner-delta} (applied twice for the d'Alembert operator in the wave equation) to convert the wave equation into tangent space $\tspace{x}{M}$,
\begin{equation}
	\left(\Box_y^2-f\beta \cdot \p_y + m^2\right)\phi(x,y)=0\qq{at}y^\mu=0\,. 
\end{equation} 
We then extend the validity of this equation to the whole tangent space $\tspace{x}{M}$, but keeping the coefficients $m$, $\beta$, and $f$, fixed at $\mcP$, 
\begin{equation}\label{eq:wave-y}
	\left[\Box_y^2-f(x)\beta(x) \cdot \p_y + m^2(x)\right]\phi(x,y)=0\,. 
\end{equation} 
The inverse Wigner transform $\phi(x)$, cf.\ Eq.\ \eqref{eq:wigner-delta}, of the solution $\phi(x,y)$ to this equation is a solution of the original wave equation \eqref{eq:wave} at point $\mcP$. 
As a linear partial differential equation with constant coefficients, Eq.\ \eqref{eq:wave-y} can be solved via Fourier transformation to the cotangent bundle,
\begin{equation}
	\phi(x,y) = \int_k \,e^{-ik\cdot y}\phi(x,k)\,,
\end{equation}
cf.\ Eq.\ \eqref{eq:F_inverse}, which then implies
\begin{equation}\label{eq:ih-wave-dispersion}
	k^2 - if(x) \beta(x) \cdot k - m^2 (x)= 0\,. 
\end{equation}
The solutions of this equation define the dispersion relations of $\phi(x,y)$. 
In order to solve Eq.\ \eqref{eq:wave-y} in similar way as in Sec.\ \ref{sec:wave-hom}, we need a foliation of tangent space in terms of spacelike hypersurfaces (with timelike normal vectors).

Let $n^\mu(x)$ be a timelike vector field, which at point $\mcP$ maps to a vector in tangent space and is normalized as $n(x) \cdot n(x) = 1$. 
We assume that this vector points into the future direction. 
At point $\mcP$, there exists an inertial frame which moves with a four-velocity $n^\mu(x)$. 
We refer to this frame as the frame of the local inertial observer. 
The vector $n^\mu(x)$ defines a foliation of tangent space $\tspace{x}{M}$ in terms of spacelike hypersurfaces, all with the same normal vector $n^\mu(x)$. 
The timelike component of an element $y^\mu$ of tangent space $\tspace{x}{M}$ is then $n\cdot y$. 
Thus any element $y^\mu$ of a spacelike hypersurface in tangent space fulfills $n\cdot y =0$.  
The cotangent space $\cspace{x}{M}$ is foliated accordingly, with $n\cdot k$ being the timelike component of a covector $k_\mu$.
If we choose $n^\mu(x)=u^\mu(x)$, the local inertial observer's frame corresponds to the local rest frame (LRF) of the fluid at each point $\mcP$. 
On the other hand, we might choose a vector field such that at every point $\mcP$ we have $n^\mu (x) = \left(1,\vb{0}\right)$ \emph{in local Minkowski coordinates}.
This choice is the local analog of the usual global noncomoving frame, in which a linear stability analysis for homogeneous equilibrium configurations is performed. 
Note, however, that in the case of inhomogeneous equilibrium configurations there is no such global frame, which necessitates the generalization to a space-time dependent $n^\mu(x)$ and the extension to the tangent space in order to perform the linear stability analysis.
For further use, we call this choice the coordinate frame (CF).
Any other choice for $n^\mu(x)$ is, of course, also possible.

 With the above considerations, we find from Eq.\ \eqref{eq:ih-wave-dispersion}, similarly as from Eq.\ \eqref{eq:wave-dispersion}, the dispersion relations $\omega_{\pm}(x,k_\perp)$, with $k_\perp^\mu \equiv (g^{\mu \nu} - n^\mu n^\nu) k_\nu$ being the components of $k^\mu$ orthogonal to $n^\mu$.
 Since the characteristic equations are covariant, one might solve them for $u\cdot k$, and then perform a Lorentz boost at $\mcP$, to find $\omega=n\cdot k$, if required.
 Summing over the two modes arising from the roots of Eq.\ \eqref{eq:ih-wave-dispersion}, and integrating over $k$, we obtain the Wigner transform $\phi (x,y)$ of the solution $\phi (x)$ to the wave equation as
\begin{equation}\label{eq:winger-wave}
	\phi (x,y) = \int_k \sum_{a=\pm} \phi_a(x,k) \delta(n\cdot k - \omega_a)e^{-ik\cdot y}\,.
\end{equation}
%The Wigner transform $\phi (x,y)$ must be square integrable on spacelike hypersurfaces in the tangent space, i.e., 
%\begin{equation}\label{eq:wave-norm-tb}
%\norm{\phi(x)}_{\tspace{x}{M}}^2 \equiv %\int\dd[4]{y}\sqrt{-g} \, %\abs{\phi(x,y)}^2 \delta(n\cdot y) < %\infty\,. 
%\end{equation}
According to Eq.\ \eqref{eq:useful_id}, the solution $\phi(x)$ to the original wave equation arises from Eq.\ \eqref{eq:winger-wave} as
\begin{equation}\label{eq:ih-wave-sol}
	\phi (x) = \int_k \sum_{a=\pm}  \phi_a(x,k)\ \delta(n\cdot k - \omega_a)\,.
\end{equation}
Note that there is no longer an exponential factor which can tell us whether a mode $\omega_a(x,k_\perp)$ is exponentially growing or not. 
Nevertheless, this information is still contained in Eq.\ \eqref{eq:ih-wave-sol}, as we will show next.

Equation \eqref{eq:winger-wave} implies that $\phi(x,y)=\sum_a \phi_a(x,y)$, where $\phi_a(x,y)$ is the Wigner transform of $\phi_a(x)$. 
Therefore, $\phi_a(x,k)\delta(n\cdot k-\omega_a)$, which according to Eq.\ \eqref{eq:winger-wave} is the Fourier transform of $\phi_a(x,y)$, fulfills Eq.\ \eqref{eq:chlift-curved}
\begin{equation}\label{eq:amp-evolution}
    \chlift_\mu \left[ \phi_a(x,k) \delta(n\cdot k-\omega_a)\right]= - ik_\mu \phi_a(x,k) \delta(n\cdot k-\omega_a)\,,
\end{equation}
where curvature terms are neglected.
We can rewrite Eq.\ \eqref{eq:amp-evolution} as
\begin{eqnarray}
    \left[\chlift_\mu  \phi_a(x,k) \right]\delta(n\cdot k-\omega_a) &=& -  \phi_a(x,k) \chlift_\mu(n\cdot k-\omega_a)
    n \cdot \p_k
    \delta(n\cdot k-\omega_a)- ik_\mu \phi_a(x,k) \delta(n\cdot k-\omega_a)
    \nn\\
    &=& n \cdot \p_k\left[ \phi_a(x,k) \chlift_\mu(n\cdot k-\omega_a)\right]
    \delta(n\cdot k-\omega_a)- ik_\mu \phi_a(x,k) \delta(n\cdot k-\omega_a)\,,
\end{eqnarray}
where we have performed an integration by parts from the first to the second line, using the fact that $n \cdot \p_k$ corresponds to $\dd / \dd k_0$ under the integral.
Using $[\chlift_\mu,\p^\nu_k]=0$, cf.\ Eq.\ \eqref{eq:chlift-comm-dk}, and $n^\nu \chlift_\mu n_\nu =  n^\nu \nabla_\mu n_\nu = 0$, we expand the first term on the right-hand side to obtain
\begin{align}
    \left[\chlift_\mu  \phi_a(x,k) \right]\delta(n\cdot k-\omega_a)   &= \left\{ \Big[n \cdot\p_k \phi_a(x,k)\Big] \chlift_\mu(n\cdot k-\omega_a) 
    - \phi_a(x,k) n_\nu\chlift_\mu\p^\nu_k\omega_a 
    % \nn\\&&
    - ik_\mu \phi_a(x,k)\right\} \delta(n\cdot k-\omega_a)
    \nn\\
    &= \left\{ \Big[ n \cdot \p_k \phi_a(x,k) \Big] \chlift_\mu(n\cdot k-\omega_a) 
    + \phi_a(x,k) [\p^\nu_k\omega_a]  \chlift_\mu n_\nu
       - ik_\mu \phi_a(x,k)\right\} \delta(n\cdot k-\omega_a)\,,
\end{align}
where we have used the fact that $\omega_a$ depends only on $x^\mu$ and the projection of $k^\mu$ orthogonal to $n^\mu$, i.e., $n \cdot \p_k\omega_a = 0$.
Finally, we use $\chlift_\nu k_\rho = 0$, cf.\ Eq.\ \eqref{eq:chlift-k}, to find
\begin{align}\label{eq:amp-master-eq-ex}
    \left[ \chlift_\mu  \phi_a(x,k) \right] \delta(n\cdot k-\omega_a)  
    &= \left\{ \Big[n \cdot\p_k \phi_a(x,k) \Big]  (k_\rho\chlift_\mu n^\rho-\chlift_\mu\omega_a) 
    + \phi_a(x,k)[\p^\nu_k\omega_a  ] \chlift_\mu n_\nu
       - ik_\mu \phi_a(x,k)\right\} \delta(n\cdot k-\omega_a)\,.
\end{align}

Let us now consider a curve $\mcC$ passing through $\mcP$, of which $n^\mu(x)$ is the tangent vector and which is parameterized with the affine parameter $\propt$,  with $\propt=0$ at $\mcP$. 
An infinitesimal change in this parameter is given by $\dd{\propt} \equiv n_\mu \dd x^\mu$.
At each point, $\propt$ can be chosen to coincide with the corresponding local inertial observer's proper time.
Since the derivative of a quantity with respect to $\propt$ is the component of the gradient of that quantity in $n^\mu$-direction,
\begin{equation}
    \dv{{}}{\propt} \equiv 
    n (x) \cdot \chlift \,,
\end{equation}
we obtain from Eq.\ \eqref{eq:amp-master-eq-ex} by contraction with $n^\mu$
\begin{align} \label{eq:amp-master-eq}
    \dv{\phi_a(x,k)}{\propt} \delta(n\cdot k-\omega_a)  &= 
    \left\{
    \Big[ n \cdot\p_k \phi_a(x,k) \Big] \left(k\cdot \dv{n}{\propt}-\dv{\omega_a}{\propt}\right)
    + \phi_a(x,k)\dv{n}{\propt}\cdot \p_k\omega_a -i\omega_a \phi_a(x,k)
       \right\}\delta(n\cdot k-\omega_a)\,.
\end{align}
The right-hand side of Eq.\ \eqref{eq:amp-master-eq} shows that, along the curve $\mcC$, the evolution of $\phi_a(x,k)$ is only partially governed by the local frequency $\omega_a(x,{k_\perp})$, as there are additional nontrivial contributions.
In the LRF, where $n^\mu(x) = u^\mu (x)$, there is a term proportional to $\dd n^\mu/ \dd \propt \equiv a^\mu$, i.e., the acceleration of the fluid along $\mcC$.
On the other hand, in the CF frame, where $n^\mu = (1,\vb{0})$, the acceleration vanishes, but the frequency still changes along $\mcC$, and there is a term proportional to $-\dd \omega_a/\dd \propt$.

We now define the norm
\begin{equation}\label{eq:physical-norm}
    \norm{\phi(\propt)}^2 = \int_{\Sigma_n(\propt)} \dd{\Sigma_n} \abs{\phi(x)}^2\,,
\end{equation}
where $\dd \Sigma_n \equiv \epsilon_{\alpha \beta \gamma \delta} \, n^\alpha \dd x^\beta \dd x^\gamma \dd x^\delta$ is the infinitesimal 3-dimensional volume element on a spacelike hypersurface $\Sigma(\propt)$ with timelike normal vector $n^\mu(x)$.
As we will show below, this norm will grow beyond bounds as $\propt \rightarrow \infty$ if there is an instability.
We can convince ourselves that this works in the case of a homogeneous equilibrium configuration and $n^\mu = (1, \vb{0})$.
Then, $\chlift_\mu$ in Eq.\ \eqref{eq:amp-master-eq-ex} reduces to $\partial_\mu$ in Minkowski coordinates, $\propt \equiv t$ up to some arbitrary constant, and 
the first two terms on the right-hand side of Eq.\ \eqref{eq:amp-master-eq-ex} vanish since $n^\mu$ and $\omega_a$ are constant in space-time. The solution of Eq.\ \eqref{eq:amp-master-eq-ex} is then simply given by
\begin{equation}
    \phi_a(x,k) = e^{-i\omega_a t+i\vb{k}\cdot \vb{x}}\phi_a(0,k)\,,
\end{equation}
where $\phi_a(0,k)$ is determined by the initial condition. 
Inserting this into Eq.\ \eqref{eq:ih-wave-sol} and the result into Eq.\ \eqref{eq:physical-norm}, we obtain after repeating similar steps as in Sec.\ \ref{sec:lin-stab-rev} an expression analogous to Eq.\ \eqref{eq:norm_unstable}.

On the other hand, if the configuration is inhomogeneous, the solution of Eq.\ \eqref{eq:amp-master-eq-ex} is not just a simple exponential factor, due to the additional terms on the right-hand side.
However, there might still exist directions in space-time for which such a solution arises.
In the next subsection, we identify these directions, which we refer to as \textit{equilibrium-preserving directions}.
After that, in Sec.\ \ref{sec:restrict}, we argue that, in the short-wavelength regime, if $\Im\omega_a(x,k_\perp) > 0$ in a subdomain of equilibrium-preserving components of $k_\mu$, the theory becomes linearly unstable.

\subsection{Equilibrium-preserving directions in tangent space}

The Wigner transform of the $\beta$-vector reads
\begin{equation} \label{eq:Wigner_beta}
\beta_\mu (x,y) = e^{ y \cdot \hlift} \beta_\mu (x)\,.
\end{equation}
Expanding the exponential, the next-to-leading order in the above equation features $y^\nu \nabla_\nu \, \beta_\mu = y^\nu \varpi_{\mu\nu}$, where we used the Killing condition \eqref{eq:killing}.
The next-to-next-to-leading order is then proportional to $y^\lambda y^\nu\nabla_\lambda \varpi_{\mu\nu} = {y^\lambda} y^\nu R_{\mu\nu\lambda\sigma}\beta^\sigma$, cf.\ Appendix\ \ref{app:voriticty}, which vanishes in flat space-time.
The same is true for all higher orders, therefore, in flat space-time
\begin{equation} \label{eq:beta_exp}
	\beta_\mu (x,y) = \beta_\mu (x) + y^\nu\varpi_{\mu\nu}(x)\,.
\end{equation}
If we compare the above with the standard relation for the $\beta$-vector in terms of the thermal vorticity in Minkowski space-time, see, e.g., Ref.\ \cite{DeGroot:1980dk},
\begin{equation}
    \beta_\mu(x) = b_\mu + x^\nu\varpi_{\mu\nu}\,, 
\end{equation}
and setting $b_\mu \equiv \beta_\mu (0)$, we find that the Wigner transform (\ref{eq:Wigner_beta}) translates the $\beta$-vector by $y^\mu$ in flat space-time,
\begin{equation}
    \beta_\mu(x+y) = \beta_\mu(x,y)\,.
\end{equation}

The directions in $\tspace{x}{M}$ for which the Wigner transform does not modify the $\beta$-vector, the so-called \textit{equilibrium-preserving directions} in $\tspace{x}{M}$, are now given by the condition
\begin{equation}\label{eq:ypara-def}
	\beta_\mu (x, y_\eqp) = e^{y_\eqp\cdot\hlift} \beta_\mu(x) \stackrel{!}{=} \beta_\mu (x)\,,
\end{equation}
where the subscript ``$\eqp$'' denotes ``equilibrium-preserving''.
Comparing Eqs.\ (\ref{eq:beta_exp}) and (\ref{eq:ypara-def}) the equilibrium-preserving directions $y_\eqp$ in flat space-time are given by the condition 
\begin{equation}\label{eq:ep-y}
	y_\eqp^\mu\varpi_{\mu\nu} (x)\stackrel{!}{=}0\,.
\end{equation}
In the accelerating configuration (without rotation, $\omega^\mu =0$), this requires that $y_\eqp \cdot u = y_\eqp \cdot \ell =0$, cf.\ Eq.\ \eqref{eq:vorticity}.
Consequently, $y_\eqp^\mu$ has only two independent components.
From Eqs.\ \eqref{eq:umu_acc} and \eqref{eq:lmu_acc} we then deduce that (in Rindler coordinates)
\begin{equation}
y_\eqp^\mu = (0, y^1, y^2, 0)\,,    
\end{equation}
i.e., the independent components are the $x$- and $y$-coordinates transverse to the direction of acceleration.

In the rotating configuration we expand $y_\eqp^\mu$ in the tetrad $(u, \ell, \psi, \zeta)$, 
\begin{equation}
    y^\mu_\eqp = y_u u^\mu + y_\ell \ell^\mu + y_\psi \psi^\mu + y_\zeta \zeta^\mu\,,
\end{equation}
as well as
$\varpi_{\mu \nu}$ according to Eq.\ \eqref{eq:vorticity}, and insert this into Eq.\ \eqref{eq:ep-y}. 
With $a^\mu \equiv a \ell^\mu$ and Eqs.\ \eqref{eq:decomp_vort} and \eqref{eq:def-zeta}, this results in
\begin{equation} \label{eq:cond}
    a y_\ell u_\nu
    +\left(a y_u - \omega_\perp y_\zeta\right) \ell_\nu
    + y_\zeta \omega_\ell \psi_\nu
    + \left(y_\ell \omega_\perp - y_\psi \omega_\ell\right) \zeta_\nu = 0\,.
\end{equation}
Since $a \neq 0$, $\omega_\perp \neq 0$ (otherwise we could not have defined the tetrad $(u,\ell, \psi, \zeta)$), we immediately deduce from Eq.\ \eqref{eq:cond} that for $\omega_\ell \neq 0$ all components of $y_\eqp^\mu$ must vanish, or in other words, an equilibrium-preserving subspace of $\tspace{x}{M}$ exists only if $\omega_\ell = 0$. 
Consequently, for $\omega_\ell =0$ we deduce from Eq.\ \eqref{eq:cond} that $y_\ell = 0$ and
\begin{equation}
    y^\mu_\eqp =  y_u u^\mu + y_\psi \psi^\mu + \frac{a}{\omega_\perp} y_u \, \zeta^\mu\,,
\end{equation}
i.e., we again have only two independent components.
With Eqs.\ \eqref{eq:umu_rot} -- \eqref{eq:decomp_vort} and \eqref{eq:zetamu_rot} we then deduce that (in cylindrical coordinates)
\begin{equation}
    y_\eqp^\mu = \left(y^0, 0, 0,  y^3\right)\,,
\end{equation}
i.e., the independent coordinates are the time coordinate and the coordinate along the direction of the rotation vector $\omega^\mu$.

In the above, we restricted the discussion to flat space-time. 
In this case, the base manifold has the same equilibrium-preserving directions as the tangent space. 
Assuming Minkowski coordinates, the components of $y^\mu$ can be considered as the coordinates of a coordinate system with origin in $x$. 
In what follows, we use the terms ``equilibrium-preserving directions'' and ``equilibrium-nonpreserving directions'' both in the base manifold and in tangent space.

\subsection{Linear-stability analysis in equilibrium-preserving directions}
\label{sec:restrict}

Now we are in a position to understand the relationship between the dispersion relations and linear stability in inhomogeneous equilibrium configurations for which equilibrium-preserving directions exist. 
As above, the equilibrium-preserving directions will be denoted by an index ``e'', while the equilibrium-nonpreserving directions will carry an index ``ne'', such that $x_\mu \equiv x^\eqp_\mu + x^\neqp_\mu$.

We now study the solutions of Eq.\ \eqref{eq:amp-master-eq}, which also fulfill Eq.\ \eqref{eq:amp-master-eq-ex}.
Note that it is the former equation whose solution is constrained by the dispersion relations arising from the wave equation \eqref{eq:wave}.
Let us now consider Eq.\ \eqref{eq:amp-master-eq} in the LRF, i.e., for $n^\mu = u^\mu$, with the following Ansatz for $\phi_a$,
 \begin{equation}\label{eq:ih_ansatz}
    \phi_a(x,k) = e^{\Gamma_{a}(\propt,x_\perp^\neqp,k_\perp)-ik^{\eqp}_{\perp} \cdot x^{\eqp}_{\perp}} \psi_a(\propt,x_\perp^\neqp,k)\,,
\end{equation}
where $k^{\eqp}_{\perp,\mu}$ is found from the condition
\begin{equation}\label{eq:restrict-k}
k^{\eqp, \mu}_{\perp} \varpi_{\mu\nu} = 0\,,
\end{equation}
and
\begin{equation}\label{eq:general-gamma}
	\Gamma_a(\propt, x_\perp^\neqp,k_\perp) = -i\int_0^{\propt} \dd{\propt'}\omega_a(\propt',x_\perp^\neqp,k_\perp)\,.
 \end{equation}
In the equilibrium-preserving directions $x^\eqp_\perp$ of flat space-time, we have $x^{\eqp}_\perp \cdot \chlift \, n_\nu = 0 = x^{\eqp}_\perp \cdot \chlift \, \omega_a$.
The first equality can be shown by using $n_\nu = u_\nu = T \beta_\nu$ and the fact that $\beta_\nu$ is a Killing vector.
The second equality arises because the only dependence of $\omega_a$ on an equilibrium-preserving direction can be through $\propt$, which is, however, orthogonal to $x_\perp^\eqp$.
Thus, Eq.\ \eqref{eq:amp-master-eq-ex} reduces to
\begin{equation} \label{eq:wave_eqp}
    \p_{\mu }^\eqp \phi_a(x,k) = - ik_{\mu }^\eqp \phi_a(x,k)\,.
\end{equation}
Projecting this equation onto the spacelike directions orthogonal to $n^\mu$, we find that the Ansatz \eqref{eq:ih_ansatz} fulfills this equation.
Plugging the Ansatz \eqref{eq:ih_ansatz} into Eq.\ \eqref{eq:amp-master-eq}, we find with Eq.\ \eqref{eq:general-gamma}
\begin{equation} \label{eq:amp-master-red}
    \dv{\psi_a(\propt,x_\perp^\neqp,k)}{\propt} \delta(n\cdot k-\omega_a)  = 
    \left\{ \Big[
    n \cdot\p_k \psi_a(\propt,x_\perp^\neqp,k) \Big]\left(k\cdot \dv{n}{\propt}-\dv{\omega_a}{\propt}\right)
    + \psi_a(\propt,x_\perp^\neqp,k)\dv{n}{\propt}\cdot \p_k\omega_a 
       \right]\}\delta(n\cdot k-\omega_a)\,.
\end{equation}
Here, the part of the Ansatz \eqref{eq:ih_ansatz} $\sim e^{-i k^{\eqp}_{\perp} \cdot x^{\eqp}_{\perp}}$ factors out immediately, since its momentum dependence is orthogonal to $n^\mu$.
Furthermore, the term $\sim -i \omega_a \phi_a(x,k)$ cancels between left- and right-hand sides.
Finally, $\Gamma_a(\propt,x_\perp^\neqp,k_\perp)$ does not depend on the components of $k^\mu$ in the direction of $n^\mu$.

We note that the terms on the right-hand side of Eq.\ \eqref{eq:amp-master-red} arise from terms in Eq.\ \eqref{eq:amp-master-eq-ex} which are proportional to $\chlift_\mu n_\nu \equiv \nabla_\mu u_\nu \sim T\varpi_{\mu\nu}$.
We remind ourselves of the discussion in Sec.\ \ref{sec:eq-cats}, namely that in an inhomogeneous equilibrium configuration, the requirement  $\beta\cdot\beta>0$ demands the existence of some boundary condition, which then introduces a characteristic length scale $\ell_{\rm vort}$ for the system.
For the pure accelerating configuration \eqref{eq:acc-conf}, this scale is $1/a_0$, while for the rigidly rotating configuration \eqref{eq:rigid-rotation} it is $1/\Omega_0$; and in both cases $T\varpi_{\mu\nu}\sim \ell_{\rm vort}^{-1}$.
Consequently, we find 
\begin{equation}\label{eq:ell-vort}
    \dv{\psi_a(\propt,x_\perp^\neqp,k)}{\propt} \sim \ell_{\rm vort}^{-1} \psi_a(\propt,x_\perp^\neqp,k)\,.
\end{equation}

Next, we insert the Ansatz \eqref{eq:ih_ansatz} into Eq.\ \eqref{eq:ih-wave-sol} and trivially perform the integration over $n\cdot k$ using the delta-function. 
Then, we decompose $k_{\perp} = k_{\perp}^{\eqp} + k_{\perp}^{\neqp}$, formally Taylor-expand $\Gamma_a(\propt, x_\perp^\neqp,k_\perp)$ in $k_{\perp}^{\neqp}$, and absorb any term beyond $k_{\perp}^{\neqp}=0$ into $\psi_a(\propt,x_\perp^\neqp,{\omega_a,k_\perp})$.
After taking the integration over $k^{\neqp}_\perp$ in Eq.\ \eqref{eq:ih-wave-sol}, we find 
\begin{equation}
\label{eq:rest-wave-sol}
	\phi(x) = \int\frac{\dd[d]{k^\eqp_\perp}}{(2\pi)^d} \sum_{a=\pm} e^{\Gamma_{a}(\propt, x_\perp^{\neqp},k_\perp^{\eqp})-ik^{\eqp}_{\perp} \cdot x^{\eqp}_{\perp}}   \Psi_a(\propt,x_\perp^\neqp, k_\perp^\eqp)\,,
\end{equation}
where $d$ is the number of spacelike equilibrium-preserving directions, and we defined
\begin{equation}
\Psi_a (\propt,x_\perp^\neqp, k_\perp^\eqp)\equiv
\int \frac{\dd[3-d]{k^\neqp_\perp}}{(2 \pi)^{
{4}-d}} \, \psi_a (\propt,x_\perp^\neqp,{\omega_a}, k_\perp)\,.
\end{equation}
Note that, on account of Eq.\ \eqref{eq:ell-vort}, we also have
\begin{equation}
\label{eq:ell-vort_2}
    \dv{\Psi_a(\propt,x_\perp^\neqp,k_\perp^\eqp)}{\propt} \sim \ell_{\rm vort}^{-1} \Psi_a(\propt,x_\perp^\neqp,k_\perp^\eqp)\,.
\end{equation}

Plugging Eq.\ \eqref{eq:rest-wave-sol} into the wave equation \eqref{eq:wave} and using the characteristic equation \eqref{eq:ih-wave-dispersion} for $k_\perp^\neqp=0$, we find a differential equation that can be solved to find $\Psi_a (\propt,x_\perp^\neqp,k_\perp^\eqp)$.
However, the functional form of $\Psi_a (\propt,x_\perp^\neqp,k_\perp^\eqp)$ is irrelevant for the following discussion; we only demand that $\phi(x)$ is square-integrable at some initial $\propt=0$ on $\Sigma_n$,
\begin{equation}
     \norm{\phi(0)}^2 < \infty\,,
\end{equation}
where $\norm{\phi(s)}^2$ is defined in Eq.\ \eqref{eq:physical-norm}. 

Now, we assume that there exists a subdomain $D_{k_\perp^\eqp}$ for which $\Im\omega_+ > 0$ for any $x$.
Then, according to Eq.\ \eqref{eq:general-gamma}, there exists a positive real-valued number $\Lambda$ such that
\begin{equation}\label{eq:local-bound}
     \Re \Gamma_+ (\propt,x_\perp^\neqp, k_\perp^\eqp) > \Lambda \propt > 0\,,
     \qq{for}
     k_\perp^\eqp \in D_{k_\perp^\eqp}\,.
 \end{equation}
 The integration in Eq.\ \eqref{eq:physical-norm} over $ x^{\eqp}_{\perp}$ yields a delta function which puts the  $k^{\eqp}_{\perp}$ of $\phi(x)$ and the corresponding $q^{\eqp}_{\perp}$ of $\phi^*(x)$ on the same value, and since we have eliminated the $k^{\neqp}_{\perp}$-dependence in $\Gamma_a(\propt,x_\perp^\neqp, k_\perp)$, after similar steps as in Sec.\ \ref{sec:lin-stab-rev} this gives rise to 
\begin{equation}
\label{eq:ih-norm-growth}
    \norm{\phi(\propt)}^2 \geq e^{2\Lambda\propt}\int \dd \Sigma_\perp^\neqp\,  F(\propt,x_\perp^\neqp)\,,
\end{equation}
where $\dd \Sigma_\perp^\neqp$ is the
$(3-d)$-dimensional hypersurface element in the equilibrium-nonpreserving directions of the 3- dimensional hypersurface element $\dd \Sigma_n$, and
\begin{equation}\label{eq:F-def}
     F(\propt,x_\perp^\neqp) = \int_{k_\perp^\eqp \in D_{k_\perp^\eqp}}\left|\Psi_+(\propt,x_\perp^\neqp,k_\perp^\eqp)+\Psi_-(\propt,x_\perp^\neqp,k_\perp^\eqp)e^{-\Delta \Gamma(\propt,x_\perp^\neqp,k_\perp^\eqp) }\right|^2\,.
\end{equation}
Here, $\Delta \Gamma \equiv \Gamma_+ - \Gamma_-$, where we have ordered the solutions $\omega_\pm$ such that $\Re\Delta \Gamma > 0$ on $D_{k_\perp^\eqp}$. 
If $F(\propt,x_\perp^\neqp)$ remains finite, the norm grows with $\propt$. 
This is sufficient for the existence of an instability.

As mentioned before, for the wave equation \eqref{eq:wave}, if $f>0$, then $\Im\omega_+ > 0$ for some subdomain $D_{k_\perp^\eqp}$. 
If $f$ is only determined by equilibrium quantities, such as the temperature, then its sign is independent of $\propt$, and the condition \eqref{eq:local-bound} is fulfilled.

In our argument, we have also assumed that $F(x^\neqp)$ in Eq.\ \eqref{eq:F-def} remains finite on the timescale $1/\Lambda$, such that the exponential factor $e^{2\Lambda\propt}$ dominates Eq.\ \eqref{eq:ih-norm-growth}.
In other words, the exponential factor $\sim e^{\Gamma_a(\propt, x_\perp^\neqp, k_\perp)}$ gives the leading behavior in Eq.\ \eqref{eq:ih_ansatz}. 
This is evidently the case in the short-wavelength regime $k\to\infty$ because the wavelength can be arbitrarily small, while $\ell_{\rm vort}$ is fixed.
Consequently, the asymptotic group velocities of waves are unaffected by a nonvanishing thermal vorticity, and therefore a theory found causal in homogeneous equilibrium configurations is also causal in inhomogeneous ones.
On the other hand, linear instabilities commonly occur in the long-wavelength regime, i.e., $k\to 0$. 
According to Eqs.\ \eqref{eq:ell-vort_2} and \eqref{eq:F-def}, the exponential factor $e^{2\Lambda\propt}$ dominates if
\begin{equation}
    \Lambda \gg \ell_{\rm vort}^{-1}\,. 
\end{equation}
 One can argue that in known applications in hydrodynamics the value of $\Lambda$, which arises from so-called nonhydrodynamic modes, is proportional to the inverse of the characteristic microscopic length scale $\ell_{\rm micro}$. 
 Therefore, if an instability occurs, it will survive if
 \begin{equation}
     \ell_{\rm vort} \gg \ell_{\rm micro}\,.
 \end{equation}
 Since $\ell_{\rm vort}$ is proportional to the size of the system, this condition is always fulfilled in the hydrodynamic regime.

\section{Application to hydrodynamics}
\label{sec:app_hyd}

In this section, we apply the ideas developed above to hydrodynamics.
We first consider the general tensor decomposition of the energy-momentum tensor with respect to the fluid four-velocity $u^\mu$ and then extend this into the cotangent space.
We note that the extension of the energy-momentum tensor into cotangent space does not commute with the tensor decomposition.
We then study as examples a perfect fluid and a dissipative fluid.
The actual stability analysis of the latter is deferred to Sec.\ \ref{sec:mis}.

\subsection{Tensor decomposition in base manifold and cotangent space}

The tensor decomposition of the energy-momentum tensor with respect to the fluid four-velocity $u^\mu$ reads
\begin{equation}
	T^{\mu\nu} = \mcE u^\mu u^\nu - \mcP \Delta^{\mu\nu} + \mcQ^\mu u^\nu + \mcQ^\nu u^\mu + \pi^{\mu\nu}\,,
\end{equation}
where the components are
\begin{equation}\label{eq:emt-comps}
	\mcE = u^\a u^\b T_{\a\b}\,,\qquad
	\mcP = -\frac{1}{3}\Delta_{\a\b} T^{\a\b}\,,\qquad
	\mcQ^\mu = \Delta^{\mu\a}u^\b  T_{\a\b}\,,\qquad
	\pi^{\mu\nu} = \Delta^{\mu\nu}_{\alpha\beta} \, T^{\a\b}\,.
\end{equation}
Following the standard procedure, we assume $T^{\mu\nu}$ to be in a state slightly out of equilibrium, $T^{\mu\nu} = T^{\mu\nu}_\eq + \d T^{\mu\nu}$ with the equilibrium energy-momentum tensor $T^{\mu\nu}_\eq$ having the perfect-fluid form,
\begin{equation}
    T^{\mu\nu}_\eq(x) = \eps_\eq(x) u^\mu_\eq(x) u^\nu_\eq(x) - p_\eq(x) \Delta^{\mu\nu}_\eq(x)\,,
\end{equation}
with $\eps_\eq(x)$ and $p_\eq(x)$ being the energy density and pressure in equilibrium, respectively, and $\Delta^{\mu \nu}_\eq(x) \equiv g^{\mu \nu} - u_\eq^\mu(x) u_\eq^\nu(x)$.
Evidently, $\mcQ^\mu_\eq$ and $\pi^{\mu\nu}_{\eq}$ vanish in equilibrium.
Consequently, up to first order in deviations from equilibrium we find
\begin{eqnarray}\label{eq:demt-base}
	\d T^{\mu\nu}(x) &=& \delta \mcE(x)  u_{\eq}^{\mu  }(x) u_{\eq}^{\nu  }(x) - \delta \mcP(x) \Delta_{\eq}^{\mu  \nu  }(x)
	+ \enp_\eq(x) \left[u_{\eq}^{\mu}(x) \delta u^{\nu}(x)   +u^\nu_\eq(x) \delta u^{\mu} (x) \right]
 \nn\\&&
 +\d\mcQ^\mu(x) u^\nu_\eq(x) + \d\mcQ^\nu (x) u^\mu_\eq(x) + \d\pi^{\mu\nu}(x)\,,
\end{eqnarray}
where $\enp_\eq(x)=\eps_\eq(x)+p_\eq(x)$ is the enthalpy density and
\begin{eqnarray}\label{eq:demt-comps}
	\d\mcE(x) &=& u^\a_\eq(x) u^\b_\eq(x) \d T_{\a\b}(x)\,,\qquad
	\d\mcP(x) = -\frac{1}{3}\Delta^{\a\b}_\eq(x) \d T_{\a\b}(x)\,,
 \nn\\
	\d\mcQ^\mu(x)  &=& \Delta_\eq^{\mu\a}(x)u^\b_\eq(x)  \d T_{\a\b}(x)-\enp_\eq(x)\d u^\mu(x)\,,\qquad
	\d \pi^{\mu\nu}(x) = \Delta^{\mu\nu}_{\eq, \alpha\beta}(x) \, \d T^{\a\b}(x)\,,
\end{eqnarray}
where $\Delta^{\mu\nu}_{\eq, \alpha\beta}(x)$ has the same form as the rank-four projection operator $\Delta^{\mu \nu}_{\a \b}$, but with the four-velocity $u$ replaced by $u_\eq$.

Since $\nabla_\mu T^{\mu \nu}_\eq =0$,  energy-momentum conservation reads $\nabla_\mu \d T^{\mu\nu}=0$.
We now extend this equation to the tangent bundle, similar to Sec.\ \ref{sec:tangent-bundle}, to obtain
\begin{equation}\label{eq:y-eom}
    \p^y_\mu \d T^{\mu\nu}(x,y)=0\,,
\end{equation}
where
\begin{equation}
    \d T^{\mu\nu}(x,y) = e^{y\cdot\hlift}
    \d T^{\mu\nu}(x)\,.
\end{equation}
This is then Fourier-transformed as
\begin{equation}\label{eq:demt-fourier}
 \d T^{\mu\nu} (x,y) = \int_k     \d T^{\mu\nu} (x,k)e^{-ik\cdot y}\,,
\end{equation}
with the EOM in the tangent bundle \eqref{eq:y-eom} giving rise to
\begin{equation}\label{eq:k-eom}
    k_\mu \d T^{\mu\nu} (x,k)  = 0\,.
\end{equation}
In order to solve this equation, similar to the wave equation in the previous section, we consider a normalized timelike vector field $n^\mu(x)$ and find the characteristic equation, the roots of which determine the dispersion relations $\omega_a(x,k_\perp)=n\cdot k$ of the modes in terms of $x$ and $k_\perp$.
We always work in the LRF, where $n^\mu \equiv u^\mu_\eq$.
We decompose $\delta T^{\mu\nu} (x,k)$ using the equilibrium four-velocity $u^\mu_\eq(x)$ as  
\begin{eqnarray}\label{eq:demt-decomp-k}
	\delta T^{\mu\nu} (x,k) &=& \delta \mcE (x,k) u_{\eq}^{\mu  }(x) u_{\eq}^{\nu  }(x) - \delta \mcP(x,k) \Delta_{\eq}^{\mu  \nu  }(x)
	+ \enp_\eq(x) \left[u_{\eq}^{\mu}(x) \delta u^{\nu}(x,k)   +u^\nu_\eq(x) \delta u^{\mu}(x,k) \right]
	\nn\\
	&& +\d\mcQ^\mu(x,k)u^\nu_\eq(x) + \d\mcQ^\nu(x,k) u^\mu_\eq(x)
    + \d\pi^{\mu\nu}(x,k)\,,
\end{eqnarray}
where
\begin{eqnarray}\label{eq:demt-comps-k}
	\d\mcE(x,k) &=& u^\a_\eq(x) u^\b_\eq(x) \d T_{\a\b}(x,k)\,,\qquad
	\d\mcP(x,k) = -\frac{1}{3}\Delta^{\a\b}_\eq(x) \d T_{\a\b}(x,k)\,,
 \nn\\
	\d\mcQ^\mu(x,k)  &=& \Delta_\eq^{\mu\a}(x)u^\b_\eq(x)  \d T_{\a\b}(x,k)-\enp_\eq(x)\d u^\mu(x,k)\,,\qquad
	\d \pi^{\mu\nu}(x,k) = \Delta^{\mu\nu}_{\eq,\alpha\beta}(x) \, \d T^{\a\b}(x,k)\,.
\end{eqnarray}
Inserting Eq.\ \eqref{eq:demt-decomp-k} into Eq.\ \eqref{eq:k-eom}, we find
\begin{eqnarray}\label{eq:fluid-eom-master}
	0 &=& \delta \mcE(x,k) u_{\eq}^{\nu  }(x)  \, k \cdot u_\eq (x)- \delta \mcP(x,k) \Delta_{\eq}^{\mu  \nu  }(x)k_\mu
	+ \enp_\eq (x)\left[\delta u^{\nu}(x,k) \, k \cdot u_\eq (x) +u^\nu_\eq (x)\, k\cdot\delta u(x,k)\right] 
	\nn\\&& 
 + u^\nu_\eq(x)\, k\cdot\d\mcQ(x,k)  + \d\mcQ^\nu(x,k)\, k\cdot u_\eq(x) + k_\mu \d\pi^{\mu\nu}(x,k)\,.
\end{eqnarray}
Let us denote the components in Eq.\ \eqref{eq:demt-comps} by $\d X^A(x)$ and the ones in Eq.\ \eqref{eq:demt-comps-k} by $\d X^A(x,k)$, where $A$ is the component index.
As in homogeneous equilibrium configurations, Eq.\ \eqref{eq:fluid-eom-master} yields a set of homogeneous linear equations of the form 
\begin{equation}\label{eq:ih-linear-set}
	M^{AB}(x,k) \d X^B(x,k) = 0\,,
\end{equation}
which has a nontrivial solution if $\det M =0$. 
This gives rise to a characteristic equation whose solutions are the dispersion relations $\omega_a = \omega_a(x,k_\perp)$. 
Consequently, according to Eq.\ \eqref{eq:winger-wave} the solution of Eq.\ \eqref{eq:y-eom} in the tangent space is found to be 
\begin{equation}\label{eq:fluid-emt-tm}
    	\d T^{\mu\nu}(x,y) = \int_k \sum_{a} \d T^{\mu\nu}(x,k)\ \delta(n\cdot k - \omega_a)\, e^{-ik\cdot y}\,.
\end{equation}
The energy-momentum tensor in the base manifold is then found as
\begin{equation}\label{eq:fluid-emt-base}
    	\d T^{\mu\nu}(x) = \int_k \sum_{a} \d T^{\mu\nu}(x,k)\ \delta(n\cdot k - \omega_a)\,,
\end{equation}
cf.\ Eq.\ \eqref{eq:ih-wave-sol}. 
We note that Eqs.\ \eqref{eq:demt-decomp-k} and \eqref{eq:demt-comps-k} look similar as Eqs.\ \eqref{eq:demt-base} and \eqref{eq:demt-comps}. 
However, integrating the quantities $\d X^A(x,k)$ over the cotangent space $\cspace{x}{M}$ does not yield the Wigner transform of the corresponding quantity $\d X^A(x)$.
As an example, let us consider $\d\mcE(x)$. 
By taking the integral over $\cspace{x}{M}$  and using Eq.\ \eqref{eq:demt-fourier}, we find
\begin{equation}
    \d\mcE(x,y) - \int_k \d \mcE (x,k)e^{-ik\cdot y} = e^{y\cdot\hlift}\left[u^\mu_\eq(x) u^\nu_\eq(x)\d T_{\mu\nu} (x)\right] -
    u^\mu_\eq(x) u^\nu_\eq(x)\d T_{\mu\nu} (x,y)
    \,,
\end{equation}
which is of order $\order{y}$
and only vanishes if $y^\mu$ is in the equilibrium-preserving directions, because then $\exp( y_\eqp \cdot \hlift) u^\mu_\eq (x)$ $=u^\mu_\eq (x)\exp( y_\eqp \cdot \hlift)$.
Consequently, only the solution $\d T^{\mu\nu}(x)$ has the form given in Eq.\ \eqref{eq:fluid-emt-base}, but not the individual components $\delta X^A(x)$, and there is an inherent freedom in defining the latter.
We will use this freedom to extend the relations between the components $\d X^A(x)$ in the base manifold to corresponding relations of the components $\d X^A(x,k)$ in $\cspace{x}{M}$.
The procedure is similar to the extension of quantities in the base manifold to the tangent bundle. 
This will be demonstrated in the following at hand of the examples of a perfect and a dissipative fluid, respectively.

\subsection{Perfect fluid}

Let us first consider a perfect fluid, for which only the components $\delta \mcE$, $\delta \mcP$, and $\delta u^\mu$ appear in the EOM \eqref{eq:fluid-eom-master}.
In the base manifold, we have $\d\mcP(x) = \vs^2(x)\d\mcE(x)$, where
\begin{equation}
    \vs^2 = \pdv{p}{\eps}\,,
\end{equation}
is the speed of sound in equilibrium. 
Using Eqs.\ \eqref{eq:demt-comps}, \eqref{eq:demt-comps-k}, and \eqref{eq:fluid-emt-base}, this implies that 
\begin{equation}
    \int_k \sum_a \left[\d \mcP(x,k)- \vs^2(x)\d\mcE(x,k)\right] \delta(n \cdot k - \omega_a) = 0\,.
\end{equation}
An obvious solution to this equation is $\d\mcP(x,k) = \vs^2(x)\d\mcE(x,k)$. 
We use the freedom in defining the components of the energy-momentum tensor in cotangent space by demanding that this relations holds everywhere in that space.
We then insert this relation into
Eq.\ \eqref{eq:fluid-eom-master} and obtain 
\begin{equation} \label{eq:perfect_fluid_modes}
	\left[\omega_a(x,k_\perp) \d\mcE(x,k) - \enp_\eq(x) k_\perp \d u_{\parallel}(x,k)\right]u^\nu_\eq(x) 
	+ \left[\enp_\eq(x) \omega_a(x,k_\perp) \d u^\nu(x,k) - v_s^2(x)\d \mcE(x,k) k^{\nu}_\perp \right]= 0\,,
\end{equation}
where  $\d u_{\parallel} (x,k) \equiv -k\cdot\delta u (x,k)/k_\perp$, $k_\perp \equiv \sqrt{-k_\perp^\a k_{\perp, \a}}$, and $k^\mu_\perp \equiv \Delta^{\mu \nu}_\eq(x) k_\nu$.
Projecting Eq.\ \eqref{eq:perfect_fluid_modes} onto $u_{\eq, \nu}(x)$ and $k_{\perp, \nu}$ results in a system of two equations of the form \eqref{eq:ih-linear-set}. 
The characteristic equation $\text{det} M^{AB}(x,k) =0$ leads to the well-known dispersion relations of the sound modes, $\omega_\pm(x,k_\perp) =\pm\vs(x) k_\perp$.

\subsection{Dissipative fluid}

As a next step, we consider a dissipative fluid. 
As will become clear in the next section, we will require derivatives of the components \eqref{eq:demt-comps} of the energy-momentum tensor.
These are computed as follows. 
Instead of $\d \mcQ(x)$ and $\d \mcQ(x,k)$ it is advantageous to introduce
\begin{subequations}\label{eq:def-tilde-Q}
\begin{align} 
    \d\tilde{\mcQ}^\mu(x)  &\equiv
    \Delta^{\mu \alpha}_{\eq} (x) \delta T_{\alpha \beta}(x) u^\beta_\eq (x) =
    \d\mcQ^\mu(x) +\enp_\eq(x)\d u^\mu(x)\,, \\
    \d\tilde{\mcQ}^\mu(x,k) & \equiv \Delta^{\mu \alpha}_{\eq} (x) \delta T_{\alpha \beta}(x,k) u^\beta_\eq (x) =  \d\mcQ^\mu(x,k) +\enp_\eq(x)\d u^\mu(x,k)\,.
\end{align}
\end{subequations}
We then take the derivative on both sides of the definitions \eqref{eq:demt-comps},
\eqref{eq:def-tilde-Q} and use Eqs.\ \eqref{eq:dwigner-delta-cb} \eqref{eq:demt-decomp-k} and \eqref{eq:demt-comps-k}, to obtain
\begin{subequations}\label{eq:emt-comps-der}
	\begin{align}
		\label{eq:mce-der}
		\nabla_\mu \d \mcE(x) &= \int_k \sum_a \left[- i k_\mu \d \mcE(x,k) - 2 T_\eq(x) \varpi_{\mu\nu}(x)\d \tilde{\mcQ} ^\nu(x,k)\right]\d (n \cdot k - \omega_a)\,,
		\\
		\label{eq:mcp-der}
		\nabla_\mu \d \mcP(x) &= \int_k \sum_a \left[-i k_\mu \d \mcP(x,k) - \frac{2}{3} T_\eq(x) \varpi_{\mu\nu}(x)\d \tilde{\mcQ} ^\nu(x,k) \right]\d (n \cdot k - \omega_a)\,,
		\\
		\label{eq:mcq-der}
\nabla_\mu \d \tilde{\mcQ}_\nu(x)  &= \int_k \sum_a \left\{-ik_\mu  \d \tilde{\mcQ}_\nu(x,k)
+ T_\eq(x) \varpi_{\mu\a}(x) \Big[u^\eq_\nu(x) \d\tilde{\mcQ}^\a(x,k)- \d\pi^{\a}_{~\,\nu}(x,k)\Big]
\right.\nn\\&\hspace*{1.6cm}  + \left.
        \Big[T_\eq(x)\varpi_{\mu\nu}(x)- a_\mu(x) u^\eq_\nu(x)\Big] \big[\d\mcE(x,k) + \d \mcP(x,k)\big]
         \Big] \right\}\d (n \cdot k - \omega_a)\,,  \\
		\label{eq:shear-der}
\nabla_\r \d \pi^{\mu\nu}(x) &= \int_k \sum_a\left\{ - i k_\r \d \pi^{\mu\nu}(x,k) + 2T_\eq(x)\varpi_{\r\a}(x)\d \pi^{\a(\mu}(x,k)u^{\nu)}_\eq(x) + 2 \Big[ T_\eq(x)\varpi_\rho^{\;(\mu}(x)- a_\r(x) u^{(\mu}_\eq(x) \Big] \d\tilde{\mcQ}^{\nu)}(x,k)
  \right. \nn \\
  & \hspace*{1.6cm} 
   -\left.\frac{2}{3}T_\eq(x) \Delta^{\mu\nu}_\eq(x)  \varpi_{\r \a}(x) \d\tilde{\mcQ}^{\a}(x,k)  \right\}\d (n \cdot k - \omega_a)\,,
	\end{align}
\end{subequations}
where we have used $\beta_{\eq, \nu}(x) \delta \tilde{Q}^\nu(x,k) =0$.
Higher-order derivatives can be computed following a similar strategy.

\subsection{Linear-stability analysis in equilibrium-preserving directions}
\label{sec:fluid-stability}

As mentioned before, the approach developed here yields the modes of the energy-momentum tensor, which are not necessarily the modes of its components \eqref{eq:emt-comps}.
However, as discussed in Sec.\ \ref{sec:restrict}, when the momenta are restricted by Eq.\ \eqref{eq:restrict-k}, the modes of the energy-momentum tensor will still provide information about whether the system is linearly stable or not.
As in Eq.\ \eqref{eq:rest-wave-sol}, we make the Ansatz
\begin{equation}
	\delta T^{\mu\nu}(x) = \int\frac{\dd[d]{{k}^\eqp_\perp}}{(2\pi)^d} \sum_{a} e^{\Gamma_{a}(\propt,x^\neqp,k^\eqp_\perp)-ik^\eqp_\perp\cdot x^\eqp_\perp}\,  \delta T^{\mu\nu}_a(\propt,x^\neqp_\perp,k^\eqp_\perp)\,.
\end{equation}
Furthermore, considering the components of the derivatives in Eqs.\ \eqref{eq:emt-comps-der} in the equilibrium-preserving directions, we find that, because of Eq.\ \eqref{eq:restrict-k} and $a_\mu(x) = T_\eq(x) \varpi_{\mu \nu}(x) u_\eq^\nu(x)$,
\begin{equation}
\nabla_{\perp, \mu}^{\eqp} \delta X^A(x)
= \int_k \sum_a \left[ - i k^{\eqp}_{\perp, \mu} \delta X^A_{{a}}(x,k) \right]\;.
\end{equation}
Therefore, the components \eqref{eq:demt-comps} can be written as 
\begin{equation}
	\delta X^A(x) = \int\frac{\dd[d]{{k}^\eqp_\perp}}{(2\pi)^d} \sum_{a} e^{\Gamma_{a}(\propt,x^\neqp,k^\eqp_\perp)-ik^\eqp_\perp\cdot x^\eqp_\perp}   \d X^A_a(\propt,x^\neqp_\perp,k^\eqp_\perp)\,,
\end{equation}
where, similar to Sec.\ \ref{sec:restrict} we have absorbed all dependence from the equilibrium-nonpreserving directions into $\d X^A_a(\propt,x^\neqp_\perp,k^\eqp_\perp)$.
We then define the norm of the components \eqref{eq:demt-comps} on spacelike hypersurfaces $\Sigma_n(\propt)$ orthogonal to $n^\mu(x)$ similar to Eq.\ \eqref{eq:physical-norm} as 
\begin{equation}\label{eq:ih-hydro-norm}
	\norm{\d X(\propt)}^2 = 
 \sum_A \int_{\Sigma_n(\propt)} \dd{\Sigma_n}\abs{\d X^A(x)}^2\,,
\end{equation}
which grows with $\propt$ if $\Im\omega_a>0$ for at least one of the modes, provided that 
\begin{equation}\label{eq:size-condition}
\ell_{\rm vort} \gg \ell_{\rm micro}\,.
\end{equation}

\section{Modes of the MIS theory in inhomogeneous equilibrium configurations}
\label{sec:mis}

In this section, we apply the approach developed in the previous section to MIS theory \cite{muller,Israel:1976tn,Israel:1979wp}.
We work in the Landau frame, where $\d \mcQ \equiv 0$. The dissipative correction \eqref{eq:demt-base} to the energy-momentum tensor thus reads with $\d\mcP = \vs^2\d\mcE+\d\Pi$,
\begin{eqnarray}\label{eq:demt-ns-bulk}
	\d T^{\mu\nu} &=& \delta \mcE  u_{\eq}^{\mu  } u_{\eq}^{\nu  } - \left(\vs^2\delta \mcE +\d \Pi\right) \Delta_{\eq}^{\mu  \nu  }
	+ \enp_\eq \left(u_{\eq}^{\mu} \delta u^{\nu}   +u^\nu_\eq \delta u^{\mu} \right) + \d \pi^{\mu\nu}\,.
\end{eqnarray}
We note that the above form is valid both in the base-manifold form of Eq.\ \eqref{eq:demt-base}, where both equilibrium quantities and perturbations are functions of $x$, as well as in the cotangent-bundle form of Eq.\ \eqref{eq:demt-decomp-k}, where equilibrium quantities are functions of $x$ and perturbations are functions of $x$ and $k$. 

The evolution of the perturbation $\d\Pi(x)$ of the bulk viscous pressure in the base manifold is given by the linearized MIS equation \cite{denicol_rischke_2022}
\begin{equation}\label{eq:mis-bulk-leom}
	\tau_\Pi u_\eq \cdot \nabla \d\Pi + \d\Pi + \zeta\nabla \cdot \d u = 0\,,
\end{equation}
where $\tau_\Pi$ is the bulk relaxation time. 
The linearized MIS EOM for the shear-stress tensor $\d \pi^{\mu \nu}(x)$ in the base manifold reads  \cite{denicol_rischke_2022}
\begin{equation}\label{eq:mis-shear-leom}
	\tau_\pi \Delta^{\mu\nu}_{\eq,\a\b} \left( u_\eq \cdot \nabla \d \pi^{\a\b} - 2 \d\pi^{\a}_\lambda \Omega^{\b\lambda}_\eq \right) + \d \pi^{\mu\nu} -2\eta\, \d\sigma^{\mu\nu} = 0\,,
\end{equation}
where $\tau_\pi$ is the shear relaxation time and $\eta$ is the shear viscosity coefficient, while $\Omega^{\mu \nu}_\eq = \half \left( \nabla^{\langle\mu\rangle}_{\phantom{\eq}} u_{\eq}^{\nu} - \nabla^{\langle\nu\rangle}_{\phantom{\eq}} u_{\eq}^{\mu}\right)$  and $\d \sigma^{\mu \nu} \equiv \Delta^{\mu \nu}_{\eq, \alpha\beta} \nabla^\a \delta u^\b$.
Note that $\Delta^{\mu \nu}_{\eq, \a \b} \nabla^\a u^\b_\eq = 0$ on account of the Killing condition \eqref{eq:killing}.
Translating Eqs.\ \eqref{eq:mis-bulk-leom} and \eqref{eq:mis-shear-leom} into cotangent space, the resulting equations, together with the energy-momentum conservation equation \eqref{eq:k-eom}, comprise a closed system that can be solved to obtain solutions of the form \eqref{eq:fluid-emt-base}.
In the following, we will explicitly demonstrate how this works.

It is advantageous to work with dimensionless quantities, i.e., we divide perturbations of the energy density $\d\mcE$, the bulk viscous pressure $\d\Pi$, and the shear-stress tensor $\d\pi^{\mu\nu}$ by the enthalpy density in equilibrium, $\enp_\eq$,
\begin{eqnarray}\label{eq:redefs}
	\delta\tilde{\mcE} \equiv \delta\mcE / \enp_\eq\,,\qquad
	\delta\tilde{\Pi} \equiv \delta\Pi / \enp_\eq\,,\qquad
	\delta\tilde{\pi}^{\mu\nu} \equiv \delta\pi^{\mu\nu}/\enp_\eq\,.
\end{eqnarray}
Next, we generalize the method proposed in Ref.\ \cite{Brito:2020nou} for the covariant decomposition of vectors and tensors into the directions of $u^\mu$, $\ell^\mu$, and directions transverse to the latter two.
To this end, it is useful to define a tetrad of four orthonormal vectors, which is different from the one defined in Sec.\ \ref{sec:eq-cats}.
The first two elements of the tetrad are $u_\eq$ and $\ell$.
To obtain the third one, we decompose the four-momentum $k^\mu$ as
\begin{equation} \label{eq:decomp_k} 
	k^\mu = T_\eq\left(\Omega u^\mu_\eq + \kappa_\ell \ell^\mu + \kappa^\mu\right)\,,
\end{equation}
where $\Omega \equiv k\cdot u_\eq / T_\eq$ is the frequency scaled by the temperature in the LRF, $\kappa_\ell \equiv -k\cdot \ell /T_\eq$, and 
\begin{equation}
	\kappa^\mu \equiv \frac{1}{T_\eq} \, \Xi^{\mu\nu} k_\nu\,,\qq{with}
	\Xi^{\mu\nu} \equiv \Delta^{\mu\nu}_\eq + \ell^\mu \ell^\nu \,.
\end{equation}
Consequently, $\tilde{\kappa}^\mu \equiv\kappa^\mu/\kappa$, with
	$\kappa \equiv \sqrt{-\kappa \cdot \kappa}$, which is orthogonal to both $u_\eq$ and $\ell$, is the third element of the tetrad.
Since we assume that $\ell$ is nonzero, the fourth element of the tetrad is found to be
\begin{equation}\label{eq:def-chi}
	\chi^\mu \equiv  \epsilon^{\mu\nu\a\b} u^\eq_\nu \ell_\a \tilde{\kappa}_\b\,.
\end{equation}
Tensors of arbitrary rank can be decomposed in terms of the tetrad $\{u_\eq,\ell,\tilde{\kappa},\chi\}$.
To begin, $\d u^\mu$ is decomposed as
\begin{equation} \label{eq:delta_u_decomp}
    \d u^\mu = \d u_\ell \ell^\mu + \d u_\kappa\tilde{\kappa}^\mu + \d u_\chi \chi^\mu\,,
\end{equation}
where
\begin{equation}
	\d u_\ell = - \ell\cdot \d u\,,\qquad 
 \d u_\kappa = -\tilde{\kappa} \cdot \d u\,,\qquad 
	\d u _\chi = -\chi \cdot \d u\,.
\end{equation}
There is no component in the direction of $u_\eq$ since $\delta u^\mu$ is orthogonal to $u_\eq^\mu$.
Then, using $\d \tilde{\pi}^{\mu \nu} = \d \tilde{\pi}^{\nu \mu}$, $\d \tilde{\pi}^{\mu \nu} u_{\eq, \nu} = 0$, and $\d \tilde{\pi}^{\mu}_{\hspace*{0.1cm} \mu} = 0$, we decompose the dimensionless shear-stress tensor $\d\tilde{\pi}^{\mu\nu}$  as
\begin{eqnarray}\label{eq:shear-decomp}
	\d \tilde{\pi}^{\mu\nu} &=& \d\pi_{\ell\ell} \left(\ell^\mu \ell^\nu - \chi^\mu\chi^\nu\right)
	+
	2 \d\pi_{\ell\kappa} \ell^{(\mu}\tilde{\kappa}^{\nu)}
	+
	2 \d\pi_{\ell\chi} \ell^{(\mu}\chi^{\nu)}
 +
	\d\pi_{\kappa\kappa}
 \left(\tilde{\kappa}^\mu \tilde{\kappa}^\nu - \chi^\mu\chi^\nu\right)
	+
	2 \d\pi_{\kappa\chi} \tilde{\kappa}^{(\mu}\chi^{\nu)}\,,
\end{eqnarray}
where
\begin{eqnarray}\label{eq:shear-comps}
	\d\pi_{\ell\ell} &=& \ell_\mu \ell_\nu \d\tilde{\pi}^{\mu\nu}\,,
	\quad
	\d\pi_{\ell\kappa}  =\ell_\mu\tilde{\kappa}_\nu  \d\tilde{\pi}^{\mu\nu}\,,
	\quad
	\d\pi_{\ell\chi} = \ell_\mu \chi_\nu \d\tilde{\pi}^{\mu\nu}\,,
	% \nonumber\\
 \quad 
	\d\pi_{\kappa\kappa} 
 % &=& 
 =
 \tilde{\kappa}_\mu \tilde{\kappa}_\nu \d\tilde{\pi}^{\mu\nu}\,,
	\quad
	\d\pi_{\kappa \chi}  = \tilde{\kappa}_\mu \chi_\nu \d\tilde{\pi}^{\mu\nu}\,.
\end{eqnarray}

Now, following the procedure explained in the previous section, we insert $\d T^{\mu\nu}(x,k)$ in the decomposed form of Eq.\ \eqref{eq:demt-ns-bulk} into the EOM \eqref{eq:k-eom} in the cotangent space, use Eqs.\ \eqref{eq:redefs}, \eqref{eq:decomp_k}, \eqref{eq:delta_u_decomp}, and \eqref{eq:shear-decomp}, and contract it with successive elements of the set $\{ u_\eq,\ell,\tilde{\kappa},\chi\}$  to find
\begin{subequations}\label{eq:fluid-eom-decomp}
	\begin{align}\label{eq:fluid-eom-u}
		  &               
		\Omega\d\tilde{\mcE}  -  \left(\kappa_\ell\d u_\ell + \kappa \d u_\kappa\right) =0\,,
		\\
		\label{eq:fluid-eom-b}
		  &
		\Omega \d u_\ell - \kappa_\ell\left(\vs^2\d \tilde{\mcE} + \d\tilde{\Pi}  + \d \pi_{\ell\ell}\right) - \kappa \d \pi_{\ell\kappa}  = 0\,,
		\\
		\label{eq:fluid-eom-k}
		  & \Omega \d u_\kappa - \kappa\left(\vs^2\d \tilde{\mcE} + \d\tilde{\Pi} + \d\pi_{\kappa\kappa} \right) - \kappa_\ell \d \pi_{\ell\kappa} = 0\,, 
		\\
		\label{eq:fluid-eom-c}
		  & \Omega \d u_\chi - \kappa_\ell \d \pi_{\ell \chi}
		- \kappa\d \pi_{\kappa \chi}= 0\,.
	\end{align}
\end{subequations}
Next, we turn to the EOM \eqref{eq:mis-bulk-leom} for the bulk viscous pressure. 
To obtain the derivative of $\d\Pi$, we set $\d\mcP=\vs^2\d\mcE+\d\Pi$ in Eq.\ \eqref{eq:mcp-der}, and use Eq.\ \eqref{eq:mce-der} to find 
\begin{align}
	\nabla_\mu \d\Pi(x) & = \int_k \sum_a \left\{- i k_\mu \d \Pi(x,k) +\left[2\vs^2(x)-\frac{2}{3}\right]T_\eq(x)\enp_\eq(x) \varpi_{\mu\nu}(x)\d u^\nu(x,k) \right. \nn \\& \hspace*{1.6cm} - \left. T_\eq(x)\pdv{\vs^2}{T}a_{\mu}(x) \d \mcE(x,k)\right\} \d (n \cdot k - \omega_a)\,.\label{eq:bulk-der}
\end{align}
Furthermore, from Eq.\ \eqref{eq:mcq-der} we find using $\d \tilde{\mcQ}^\mu = h_\eq \d u^\mu$ and the definitions \eqref{eq:redefs}
\begin{align} \label{eq:nabla_delta_u}
\nabla_\mu \d u_\nu (x) & = \int_k \sum_a \Big\{ - i k_\mu \d u_\nu(x,k) + T_\eq(x) \varpi_{\mu \a} (x) \big[ u_\nu^\eq(x) \d u^\a (x,k) - \d \tilde{\pi}^\a_{\;\nu} (x,k) \big] \nn \\
& +  \big[ T_\eq(x) \varpi_{\mu \nu}(x) - a_\mu(x) u_\nu^\eq(x) \big] \left[ \d \tilde{\mcE}(x,k) + \d \tilde{\mcP}(x,k) \right] - \left[ 1 + \frac{1}{v_s^2(x)} \right] a_\mu(x) \d u_\nu (x,k) \Big\} \d (n \cdot k - \omega_a)\,.
\end{align}
Contracting the indices, we obtain
\begin{equation}\label{eq:div-u}
\nabla \cdot \d u(x) = \int_k
\sum_a \left\{- i k \cdot \d u(x,k) - \left[2+\frac{1}{\vs^2(x)}\right]\, a (x)\cdot \d u(x,k)\right\} \d (n \cdot k - \omega_a)\,.
\end{equation}
Finally, we insert Eqs.\ \eqref{eq:bulk-der} and \eqref{eq:div-u} into Eq.\ \eqref{eq:mis-bulk-leom} and demand that the integrand vanishes on the whole cotangent space.
Using Eqs.\  \eqref{eq:redefs}, \eqref{eq:decomp_k}, and \eqref{eq:delta_u_decomp}, this gives rise to
\begin{eqnarray}\label{eq:mis-bulk-eom-sc}
	\left(1-i\bulkr \Omega\right) \d\tilde{\Pi}  + \left(\alpha\bulkrc + i\bulkc\kappa_\ell\right)\delta u_\ell + i\bulkc \kappa\delta u_\kappa &=& 0\,,
\end{eqnarray}
where we defined the quantities
\begin{equation}\label{eq:bulk-coeffs}
	\alpha \equiv \frac{a}{T_\eq}\,, \qquad \bulkr \equiv \tau_\Pi T_\eq\,,\qquad \bulkc \equiv \frac{T_\eq \zeta}{\enp_\eq}\,, 
 \qquad
	\bulkrc \equiv  \left(2 + \inv{\vs^2}\right)\bulkc - \frac{2}{3}\left(1 - 3\vs^2\right)\bulkr
	\,.
\end{equation}
We note that only the acceleration (via $\alpha$) appears in Eq.\ \eqref{eq:mis-bulk-eom-sc}, but not the kinematic vorticity. 
In other words, the bulk viscous pressure couples only to the acceleration and not directly to the rotation, as expected.

The EOM \eqref{eq:mis-shear-leom} for the shear-stress tensor requires a similar treatment.
Using Eq.\ \eqref{eq:shear-der}, we find 
\begin{equation} \label{eq:first_term_in_120}
	\Delta^{\mu \nu}_{\eq, \a \b} (x) u_\eq(x) \cdot \nabla {\delta \pi}^{\a \b}(x) = \Delta^{\mu \nu}_{\\eq, \a \b} (x) \int_k\left[- i \, k\cdot u_\eq(x)\, {\d \pi}^{\a\b}(x,k)- 2 \enp_\eq(x) a^{\a}(x)\d u^{\b}(x,k)\right]\,.
\end{equation}
From Eq.\ \eqref{eq:nabla_delta_u} one readily computes $\d \sigma^{\mu \nu} = \Delta^{\mu \nu}_{\eq, \a \b} \nabla^\a \d u^\b$.
Plugging the result and Eq.\ \eqref{eq:first_term_in_120} into Eq.\ \eqref{eq:mis-shear-leom}, and using Eqs.\ \eqref{eq:decomp_k} and \eqref{eq:delta_u_decomp}, as well as $\alpha = a /T_\eq$, we obtain
\begin{eqnarray}\label{eq:mis-shear-eom-sc}
	0=&& \left(1-i\shr \Omega\right) \d\tilde{\pi}^{\mu\nu}
	+2i\shc\left[\kappa_\ell \ell^{(\mu}\d u^{\nu)}+\kappa^{(\mu}\d u^{\nu)}+\frac{1}{3}\left(\kappa_\ell\d u_\ell + \kappa \d u_\kappa\right)\Delta^{\mu\nu}_\eq\right]\nn\\&&
 + 6\alpha \shrc\left(\ell^{(\mu}\d u^{\nu)} +  \frac{1}{3}\d u_\ell \Delta^{\mu\nu}_\eq\right) 
 -2\frac{\shr+\shc}{T_\eq}\d\tilde{\pi}_\a^{~(\mu}\Omega^{\nu)\a}_\eq \,,
\end{eqnarray}
where
\begin{equation}\label{eq:shear-coeffs}
	\shr = T_\eq \tau_\pi\,,\qquad \shc = \frac{T_\eq \eta}{\enp_\eq}\,,
    \qquad 
	\shrc = \inv{3}\left[ \left( 1 +\frac{1}{\vs^2}\right)\shc-\shr\right]\,.
\end{equation}
% {\color{red}
% I find the coefficient of $\d\pi_{\ell\ell}$ to be proportional to
% \begin{eqnarray*}
%     -\ell^\mu\tilde{\kappa}^\nu \Omega_{\mu\nu} &=& -\ell^\mu\tilde{\kappa}^\nu \epsilon_{\mu\nu\a\b}u^\a_\eq\omega^\b\\
%     &=& \omega_\perp\tilde{\kappa}^\nu \epsilon_{\nu\mu\a\b}\ell^\mu u^\a_\eq\psi^\b\\
%     &=& -\omega_\perp\tilde{\kappa}^\nu \epsilon_{\nu\a\mu\b}u^\a_\eq\ell^\mu\psi^\b\\
%     &=& -\omega_\perp\tilde{\kappa}^\nu \zeta_{\nu}\\
%     &=& + \omega_\perp \kappa_\zeta\,.
% \end{eqnarray*}
% }
Finally, we use Eq.\ \eqref{eq:shear-comps} to decompose Eq.\ \eqref{eq:mis-shear-eom-sc} into five independent equations,
\begin{subequations}\label{eq:shear-eom-decom}
	\begin{eqnarray}
		\label{eq:shear-eom-decom-ll}
		0&=& \left(1-i\shr \Omega\right)\d\pi_{\ell\ell}
        -\frac{2}{3}i\shc\kappa\d u_\kappa 
		+\frac{4}{3}\left(i\shc\kappa_\ell +3\alpha\shrc\right) \d u_\ell 
-  \frac{2\omega_\perp\left(\shc + \shr\right)}{\kappa T_\eq}
 \left(\kappa_\zeta\d\pi_{\ell\kappa}+\kappa_\psi\d\pi_{\ell\chi}\right)
		\,,
		\\
		\label{eq:shear-eom-decom-lk}
		0&=& \left(1-i\shr \Omega\right)\d\pi_{\ell\kappa}
  +i\shc {\kappa} \d u_\ell 
		+\left(i\shc\kappa_\ell +3\alpha \shrc\right)\d u_\kappa 
		-\frac{\omega_\perp\left(\shc + \shr\right)}{\kappa T_\eq}
 \left[\kappa_\zeta\left(\d\pi_{\kappa\kappa} - \d\pi_{\ell\ell}\right) + \kappa_\psi\d\pi_{\kappa\chi}\right]
		\nn\\
		&&
		+\frac{\omega_\ell\left(\shc +  \shr\right)}{T_\eq}\d \pi_{\ell\chi}
		\,,
		\\
		\label{eq:shear-eom-decom-lc}
		0&=& \left(1-i\shr \Omega\right)\d\pi_{\ell\chi} 
		+\left(i\shc \kappa_\ell +3\alpha{\shrc} \right)\d u_\chi 
	   -\frac{\omega_\perp\left(\shc+\shr\right)}{\kappa T_\eq}\left[\kappa_\zeta\d\pi_{\kappa\chi}-\kappa_\psi\left(2\d\pi_{\ell\ell}+\d\pi_{\kappa\kappa}\right)\right]
    \nn\\
    &&-\frac{\omega_\ell\left(\shc+\shr\right)}{ T_\eq}\d\pi_{\ell\kappa}
    \,,
		\\
		\label{eq:shear-eom-decom-kk}
		0&=& \left(1-i\shr \Omega\right)\d\pi_{\kappa\kappa} 
		+ \frac{4}{3}i\shc\kappa\d u_\kappa
		-\frac{2}{3}\left(i\shc \kappa_\ell +3\alpha\shrc\right)\d u_\ell 
		+\frac{2\omega_\perp\left(\shc +  \shr\right)}{\kappa T_\eq}\kappa_\zeta\d \pi_{\ell\kappa}
  \nn\\
		&& 
		+\frac{2\omega_\ell\left(\shc + \shr\right)}{T_\eq}\d \pi_{\kappa\chi}
		\,,
		\\
		\label{eq:shear-eom-decom-kc}
		0&=& \left(1-i\shr \Omega\right)\d\pi_{\kappa\chi}
		+ i\shc\kappa\d u_\chi	
        +\frac{\omega_\perp\left(\shc + \shr\right)}{\kappa T_\eq}
 \left(\kappa_\zeta\d\pi_{\ell\chi}+\kappa_\psi\d\pi_{\ell\kappa}\right)
  %  \nn\\
		% && 
		-\frac{\omega_\ell\left(\shc + \shr\right)}{T_\eq}\left(\d \pi_{\ell\ell}+2\d\pi_{\kappa\kappa}\right)
		\,,
	\end{eqnarray}
\end{subequations}
where
\begin{equation} \label{eq:def-kappa}
    \kappa_\zeta = -\kappa \cdot \zeta\,,\qquad
    \kappa_\psi = -\kappa \cdot \psi\,.
\end{equation}
Note that $\kappa = \sqrt{\kappa_\zeta^2 + \kappa_\psi^2}$. 
In order to derive Eqs.\ \eqref{eq:shear-eom-decom}, we have in particular used Eqs.\  \eqref{eq:decomp_vort}, \eqref{eq:def-zeta}, and \eqref{eq:def-chi} to obtain
\begin{equation}\label{eq:chi-to-zeta}
	\psi \cdot \chi =  \frac{\omega \cdot \chi}{\omega_\perp} = - \frac{\kappa\cdot\zeta}{\kappa}\,,
	\qquad
	\kappa\cdot\omega = \omega_\perp\kappa\cdot\psi\,.
\end{equation}
We note that both acceleration (in terms of $\alpha$) and kinematic vorticity (in terms of $\omega_\ell$ and $\omega_\perp$) appear in Eqs.\ \eqref{eq:shear-eom-decom}.

To recover the characteristic equation in the limit of a homogeneous equilibrium configuration, we first set $\kappa_\ell$, $\alpha$, $\omega_\ell$, and $\omega_\perp$ to zero in  Eqs.\ \eqref{eq:fluid-eom-decomp}, \eqref{eq:mis-bulk-eom-sc}, and \eqref{eq:shear-eom-decom}. 
Consequently, Eq.\ \eqref{eq:shear-eom-decom-lc} yields $\d\pi_{\ell\chi}=0$.
This is because, by taking the homogeneous limit, the rotation symmetry with respect to $\kappa$ is restored and the equations are symmetric under $\ell \leftrightarrow \chi$.
Using this symmetry, the fact that $\d \tilde{\pi}^{\mu \nu}$ is traceless gives the condition
$\d \pi_{\ell \ell} = - \half \d \pi_{\kappa \kappa}$, which renders the equation for $\d \pi_{\ell \ell}$ identical to that for $\d \pi_{\kappa \kappa}$. 
Ultimately, the system of equations is reduced to six equations for the six variables $\{\d\tilde{\mcE}, \d u_\kappa, \d u_\chi, \d\tilde{\Pi}, \d\pi_{\kappa\kappa},\d\pi_{\kappa\chi}\}$,
\begin{subequations}\label{eq:mis-eom-decomp-h}
\begin{eqnarray}       
		\Omega\d\tilde{\mcE}  -  \kappa \d u_\kappa &=&0\,,
		\\
		  \Omega \d u_\kappa - \kappa\left(\vs^2\d \tilde{\mcE} + \d\tilde{\Pi} + \d\pi_{\kappa\kappa} \right) &=&0\,, 
		\\
		  \Omega \d u_\chi - 
		\kappa\d \pi_{\kappa \chi}&=& 0\,,
  \\
  	\left(1-i\bulkr \Omega\right) \d\tilde{\Pi}  + i\bulkc \kappa\delta u_\kappa &=& 0\,,
		\\
		\label{eq:shear-eom-decom-kk}
		\left(1-i\shr \Omega\right)\d\pi_{\kappa\kappa} 
		+ \frac{4}{3}i\shc\kappa\d u_\kappa
  &=& 0 \,,
  \\
		 \left(1-i\shr \Omega\right)\d\pi_{\kappa\chi}
		+ i\shc\kappa\d u_\chi	
     &=& 0
		\,.
	\end{eqnarray}
\end{subequations}
Writing the above in the form \eqref{eq:linear-set} and setting $\det M =0$, we find that, as is well-known, the characteristic equation decomposes into the characteristic equations for the so-called shear and sound channels, which read
\begin{subequations}
\begin{eqnarray}\label{eq:hom-mis-disp}
	&& \left(1 - i \shr \Omega\right) \Omega + i \shc \kappa^2 = 0\,,
	\\
	&&
	 \left(1 - i \bulkr \Omega\right) \left(1 - i \shr \Omega\right) \left( \Omega^2 - \vs^2 \kappa^2\right) + i \Omega \kappa^2 \left[ \bulkc \left( 1 - i \shr \Omega\right) + \frac{4}{3} \shc \left(1 - i \bulkr \Omega\right) \right] = 0\,,
 \label{eq:hom-mis-disp2}
\end{eqnarray}
\end{subequations}
The imaginary parts $\Im\Omega_a$ of the roots of Eq.\ \eqref{eq:hom-mis-disp} are $ \leq 0$, provided the relaxation time $\shr >0$, thus implying linear stability.
Using the Routh-Hurwitz criterion, we find that the imaginary parts $\Im\Omega_a$ of the roots of Eq.\ \eqref{eq:hom-mis-disp2} are $\leq 0$ if $\shr$, $\shc$, $\bulkr$, and $\bulkc$ are positive.
These are the well-known conditions for linear stability of MIS theory in the LRF.
Taking the limit $\kappa\to\infty$ and demanding that the asymptotic group velocity does not exceed the speed of light, we find the linear causality conditions \cite{Pu:2009fj}
\begin{equation}\label{eq:mis-causality}
	\shr > \shc\,,\qquad \frac{4\shc}{3\shr} + \frac{\bulkc}{\bulkr} < 1 - \vs^2\,.
\end{equation}
One can show that these conditions, together with the stability conditions in the LRF, lead to linear stability in any frame \cite{Gavassino:2021owo}.

In inhomogeneous equilibrium configurations, Eqs.\ \eqref{eq:fluid-eom-decomp}, \eqref{eq:mis-bulk-eom-sc}, and \eqref{eq:shear-eom-decom} comprise a set of linear equations of the form \eqref{eq:ih-linear-set}, which, by setting $\det M =0$, yields the dispersion relations.
Calculating the determinant of the $(10\times 10)$ matrix $M$ is cumbersome.
The reason is that the rotational symmetry is broken and therefore one can no longer decompose the characteristic equation into shear and sound channels.
In the following, we restrict our attention to certain special cases.

\subsection{Nonzero bulk viscous pressure, zero shear-stress tensor}
\begin{figure}
	\centering
	\includegraphics[width=0.45\textwidth]{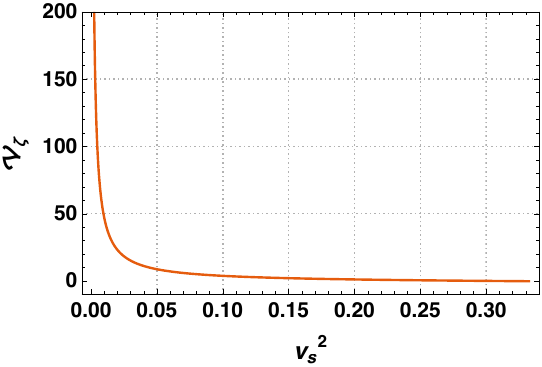}
	\caption{\amm{$\bulkrc$ as a function of $\vs^2$, with the parametrization of Eq.\ \eqref{eq:Dore_param}. For $\vs^2 \gtrsim 0.31$, $\bulkrc <0$, which is not visible in this figure.}}
	\label{fig:bulkrc}
\end{figure}
Let us first consider the case that the bulk viscous pressure is the only source of dissipation.
Consequently, the system of linearized EOMs is constituted by Eqs.\ \eqref{eq:fluid-eom-decomp}, where $\d\pi_{\mu\nu} =0$, as well as Eq.\ \eqref{eq:mis-bulk-eom-sc}.
From Eq.\ \eqref{eq:fluid-eom-c} we then find $\d u_\chi =0$, similar as for homogeneous equilibrium configurations, i.e., there is no mode transverse to both $\ell$ and $\tilde{\kappa}$.
The remaining four equations for the four variables $\{\d\tilde{\mcE},\d u_\ell,\d u_\kappa,\d \tilde{\Pi}\}$ give rise to a fourth-order characteristic equation. 
One solution is $\Omega =0$, while the other three are given by the roots of 
\begin{eqnarray}\label{eq:bulk-only-lrf}
	(1-i\bulkr\Omega) \left[ \Omega^2 - v_s^2 (\kappa^2 + \kappa_\ell^2) \right] + \Omega \left[ \alpha \bulkrc \kappa_\ell + i \bulkc (\kappa^2 + \kappa_\ell^2) \right] &=&0\,.
\end{eqnarray}
The equilibrium-preserving directions can be identified via Eq.\ \eqref{eq:restrict-k}. 
With Eqs.\ \eqref{eq:vorticity}, \eqref{eq:decomp_vort}, \eqref{eq:def-zeta}, \eqref{eq:decomp_k}, \eqref{eq:def-chi}, \eqref{eq:def-kappa}, and \eqref{eq:chi-to-zeta}, we obtain
\begin{equation} \label{eq:cond_eq_pres}
    0 = a \kappa_\ell u_\nu^\eq + 
    \omega_\perp \kappa_\zeta \ell_\nu +
    (\omega_\perp \kappa_\ell +  \omega_\ell \kappa_\psi) \zeta_\nu
    - \omega_\ell \kappa_\zeta \psi_\nu\;.
\end{equation}
Since $\{u_\eq, \ell, \zeta, \psi\}$ form an orthogonal basis, we have to demand that all coefficients vanish, leading to the requirement $\omega_\ell=\kappa_\zeta=\kappa_\ell=0$, i.e., the equilibrium-preserving direction is the $\psi$ direction, as $\kappa_\psi$ can be nonzero.
Using this in Eq.\ \eqref{eq:bulk-only-lrf}, the latter reduces to its homogeneous counterpart \eqref{eq:hom-mis-disp2} (for $\shr =  \shc =0$).
Therefore, the stability conditions found in the homogeneous equilibrium configuration in the LRF, i.e., $\bulkr >0$, $\bulkc>0$, extend to inhomogeneous equilibrium configurations.
However, the imaginary parts of the roots of Eq.\ \eqref{eq:bulk-only-lrf} can become positive in the equilibrium-nonpreserving direction $\ell$, as we will show now.  

By performing a Routh-Hurwitz analysis \cite{hastir2023generalized} on Eq.\ \eqref{eq:bulk-only-lrf} we find that $\Im\Omega_a \leq 0$ for all values of $\kappa$ and $\kappa_\ell$ only if $\bulkr > 0$, and 
\begin{subequations}\label{eq:bulk-rh}
\begin{align}
    \bulkc \kappa^2 + (\bulkc - \bulkr \alpha^2 \bulkrc^2) \kappa_\ell^2 & > 0 \,, \\
    \bulkc^2 \kappa^2 + \left[ \bulkc^2 - (\bulkc + v_s^2 \bulkr) \bulkr \alpha^2 \bulkrc^2 \right] \kappa_\ell^2 & > 0\;.
\end{align}
\end{subequations}
Therefore, for $\kappa=0$, we must have
\begin{equation}\label{eq:bulk-conditions}
	\mcC_4 \equiv \bulkc - \bulkr \alpha^2 \bulkrc^2 > 0\,,\qquad
	\mcC_6 \equiv \bulkc^2 - (\bulkc + v_s^2 \bulkr) \bulkr \alpha^2 \bulkrc^2 > 0\,.
\end{equation}
Setting $\mcC_4 = 0$, we find that for any set $\{\vs, \bulkc,\bulkr\}$, there exists a critical value
\begin{equation} \label{eq:alpha4}
\alpha^c_4 \equiv \sqrt{\frac{\bulkc}{\bulkr\bulkrc^2} }\,,
\end{equation}
 such that for $\alpha>\alpha_4^c$, $\mcC_4$ is negative and therefore $\Im\Omega_a>0$ for at least one of the modes.
A similar critical value
\begin{equation} \label{eq:alpha6}
    \alpha^c_6 \equiv
        \frac{\bulkc}{\sqrt{\bulkr\left(\bulkrc^2\bulkc+\vs^2\bulkr\right)}}\,,
\end{equation}
exists, such that for $\alpha>\alpha_6^c$, $\mcC_6$ is negative.
For positive values of $\bulkc$ and $\bulkr$, $\alpha_6^c\leq \alpha_4^c$. 
Therefore $\Im\Omega_a\leq 0$ if and only if $\alpha \leq \alpha_6^c$.

\begin{figure}
	\centering\includegraphics[width=0.45\textwidth]{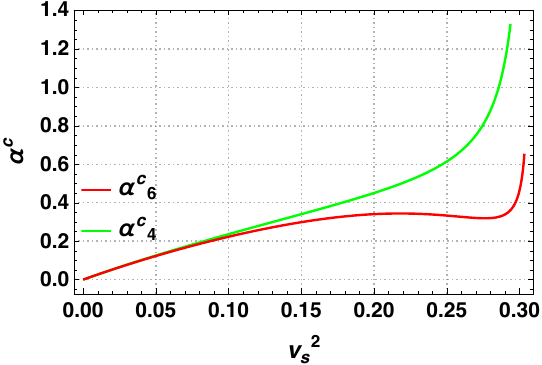}
	\caption{\amm{The critical parameters \eqref{eq:alpha4} and \eqref{eq:alpha6} as a function of $\vs^2$.}}
\label{fig:alphavsplot}
\end{figure}

Now, let us consider the following parametrization of the bulk transport coefficients which ensures linear causality for $\vs^2<1/3$  \cite{Dore:2020jye},
\begin{equation} \label{eq:Dore_param}
	\bulkc = \frac{3}{2\pi} \left(1-3 \vs^2\right)\,,\qquad\bulkr = \frac{9}{10 \pi} \left(1-3\vs^2\right)^{-1}\,.
\end{equation}
In the range $0<\vs^2<1/3$, the coefficient $\bulkrc$, cf.\ Eq.\ \eqref{eq:bulk-coeffs}, is a function of $\vs$ that, as can be seen in Fig.\ \ref{fig:bulkrc}, becomes very large for smaller values of $v_s^2$.
Consequently, the lower bounds  $\alpha^c_{4,6}$ become small, as illustrated in Fig.\ \ref{fig:alphavsplot}. 

In order to estimate the typical magnitude of $\alpha$ in applications to heavy-ion collisions, let us imagine a cylinder of QGP, which is rigidly rotating according to the configuration \eqref{eq:rigid-rotation}, and let us take $T_0=200\MeV$ and $\Omega_0=6\MeV$.
The latter number corresponds to the order of magnitude of angular velocities reported in heavy-ion collisions, i.e., $\sim 10^{22} \, \text{s}^{-1}$ \cite{STAR:2017ckg}.
Inserting these numbers into Eq.\  \eqref{eq:amu_rot} and using $\alpha\equiv \sqrt{-a\cdot a}/T_\eq$, we find $\alpha$ as a function that monotonously increases with radial distance $\rho$.
For example, it assumes the concrete values $\alpha(1\fm) \approx 0.01$, and $\alpha(5\fm) \approx 0.04$.
In order to estimate the critical values \eqref{eq:alpha4} and \eqref{eq:alpha6}, we assume that $\vs^2 = 0.2$, which is reasonable at $T_0=200 \MeV$. 
From Fig.\ \ref{fig:alphavsplot} one then reads off that the values of $\alpha$ are much smaller than the critical ones for the violation of the conditions \eqref{eq:bulk-conditions}, i.e., $\alpha^c_6 \approx 0.34$.
Thus, for these assumptions, there is no instability.
Nevertheless, even if the conditions \eqref{eq:bulk-conditions} are violated, it does not necessarily mean that there is an instability the amplitude of which grows without bounds, because the momenta of the corresponding modes point into the equilibrium-nonpreserving directions (in our case the direction of acceleration $\ell$) which were absorbed into $\d X^A_a(x^\neqp,k^\eqp_\perp)$ in the linear-stability argument of Sec.\ \ref{sec:fluid-stability}.

It is illuminating to investigate the modes arising from the roots of Eq.\ \eqref{eq:bulk-only-lrf} in the long- and short-wavelength regimes.
Let us first consider the former, for which $\kappa_t \equiv \sqrt{\kappa^2+\kappa^2_\ell} \ll 1$. 
We then expand Eq.\ \eqref{eq:bulk-only-lrf} in terms of $\kappa_t$, with $\kappa_\ell/\kappa_t$ being an arbitrary number between $-1$ and $+1$.
Solving the resulting equation order by order, we obtain two hydrodynamic sound modes and one nonhydrodynamic mode, 
\begin{subequations}\label{eq:bulk-modes}
	\begin{eqnarray}\label{eq:bulk-sound}
		\Omega_{\rm sound} &=& \pm\sqrt{\vs^2\kappa_t^2 + \inv{4}\alpha^2\bulkrc^2\kappa_\ell^2} - \inv{2}\alpha\bulkrc\kappa_\ell
		\\\nn&&
		-\frac{i}{2}\left(1\mp\frac{\alpha\bulkrc\kappa_\ell}{\sqrt{4\vs^2\kappa_t^2 + \alpha^2\bulkrc^2\kappa_\ell^2}}\right)\left\{\bulkc\kappa_t^2+\bulkr\left[\vs^2\kappa_t^2-\left(\sqrt{\vs^2\kappa_t^2 + \inv{4}\alpha^2\bulkrc^2\kappa_\ell^2}\mp \inv{2}\alpha\bulkrc\kappa_\ell\right)^2\right]\right\} +\order{\kappa_t^3}\,,\\
		\label{eq:bulk-gapped}
		\Omega_{\rm nonhydro} &=& -\frac{i}{\bulkr} + \alpha\bulkrc\kappa_\ell + i \left(\bulkc \kappa_t^2-\alpha^2\bulkr\bulkrc^2\kappa_\ell^2\right) +\order{\kappa_t^3}\,.
	\end{eqnarray}    
\end{subequations}
This expansion reveals the significance of $\bulkrc$. 
Letting $\kappa = 0$ in Eq.\ \eqref{eq:bulk-sound} we find the group velocity of the sound mode in the direction of acceleration to be
\begin{equation}
	\pdv{\Re\Omega_{\rm sound}}{\kappa_\ell} = \pm\sqrt{\vs^2 + \inv{4}\alpha^2\bulkrc^2}-\inv{2}\alpha\bulkrc + \cdots\,.
\end{equation}
Assuming $\alpha\ll 1$, the leading term in the group velocity is $\pm\vs-\tfrac{1}{2}\alpha\bulkrc$, i.e., that velocity is modified in the direction of acceleration. 
While the absolute value of the group velocity increases for the mode originally moving with $-\vs$, it decreases for the other one.
Thus, a nonzero acceleration breaks the symmetry of the sound waves moving in opposite directions relative to the acceleration.

Next, let us assume the short-wavelength regime, i.e.,  $\kappa_t \gg 1$.
In this limit, we find
\begin{eqnarray}\label{eq:bulk-asymp}
	\Re\Omega \sim  \pm \kappa_t\sqrt{\vs^2+\frac{\bulkc}{\bulkr}}\,.
\end{eqnarray}
This means that the asymptotic group velocity is independent of $\kappa_\ell/\kappa$ and remains smaller than the speed of light, with the same conditions that are found for the homogeneous case \eqref{eq:mis-causality}.
Furthermore, Eq.\ \eqref{eq:bulk-asymp} shows that, in the short-wavelength regime, the symmetry of the sound modes \amm{traveling} in opposite directions is recovered. 

\subsection{Conformal MIS theory}

Let us now consider a conformal fluid, for which $\vs^2=1/3$ and $\d\Pi=0$. 
Inserting this into Eqs.\ \eqref{eq:fluid-eom-decomp}, and \eqref{eq:shear-eom-decom}, we find a system of nine equations for nine variables.
The characteristic equation of this system of equations is a polynomial of order nine, which can in general not be further decomposed due to the lack of rotational symmetry in the direction orthogonal to the momentum.
The general characteristic equation is not shown here since it is too complicated, but we comment on some aspects. 

Let us first consider the characteristic equation in the long-wavelength regime.
Similarly to the previous subsection, we expand the characteristic equation in terms of $\kappa_t$, keeping the ratios $\kappa_\zeta/\kappa_t$, $\kappa_\psi/\kappa_t$, and $\kappa_\ell/\kappa_t$ arbitrary numbers between -1 and +1 (but respecting the constraints $\kappa^2 = \kappa_\zeta^2 + \kappa_\psi^2$ and $\kappa_t^2 = \kappa_\ell^2 + \kappa^2$). 
At zeroth order in $\kappa_t$, the characteristic equation has four roots with $\Omega_{1,2,3,4}=0$ and five other roots solving
\begin{equation}\label{eq:shear-om-0}
	\left(1 -i\Omega\shr\right) 
 \left[\left(1-i\Omega\shr\right)^4
 +5\left(\mcC_\ell^2+\mcC_\perp^2\right)\left(1-i\Omega\shr\right)^2
+4\left(\mcC_\ell^2+\mcC_\perp^2\right)^2\right]
 = 0 \,,
\end{equation}
where 
\begin{equation}
    \mcC_\ell \equiv \frac{\shr+\shc}{T_\eq}\omega_\ell\,,\qquad 
    \mcC_\perp \equiv \frac{\shr+\shc}{T_\eq}\omega_\perp\,.
\end{equation}
Consequently, the roots of Eq.\ \eqref{eq:shear-om-0} are five  nonhydrodynamic modes, which are distinct only if the kinematic vorticity does not vanish,
\begin{equation}\label{eq:mis-modes-0}
	\Omega_{5} = -\frac{i}{\shr}\,,\quad
	\Omega_{6,\,7} = \frac{-i\pm \sqrt{\mcC_\ell^2+\mcC_\perp^2}}{\shr} \,,\quad
	\Omega_{8,\,9} = \frac{-i\pm 2\sqrt{\mcC_\ell^2+\mcC_\perp^2}}{\shr} \,.
\end{equation}
We note that these modes differ only in their real parts.
It is interesting to note that the last four modes have a nonzero real part even for vanishing momentum.
We attribute this to the Coriolis force introduced by a nonvanishing rotation.

For the hydrodynamic modes $\Omega_{1,2,3,4}$, i.e., the ones which vanish for zero momentum, the calculation of the term which is of first order in momentum is cumbersome.
Therefore, we restrict ourselves to the equilibrium-preserving $\psi$ direction in the rigidly rotating configuration.
After setting $\omega_\ell = \kappa_\zeta=\kappa_\ell=0$, cf.\ discussion after Eq.\ \eqref{eq:cond_eq_pres}, in the first-order term of the expansion of the characteristic equation, we find two vanishing roots $\Omega_{1,2} = 0$ and two nonvanishing roots, which correspond to the sound modes and read 
\begin{equation}
    \Omega^{\eqp}_{3,4} = \pm\sqrt{\frac{1}{3} - \frac{9\alpha^2\shrc^2}{1+\mcC_\perp^2}}\; \kappa_\psi\,.
\end{equation}
One notices that, in contrast to the case with bulk viscosity only, the group velocity is modified in the equilibrium-preserving $\psi$ direction. 
The other two hydrodynamic modes are modifications of the shear modes in the homogeneous case (with dispersion relation $\Omega = - i\shc\kappa^2$), cf.\ Eq.\ \eqref{eq:hom-mis-disp}, which in the equilibrium-preserving $\psi$ direction have a contribution of the form $-i\shc\kappa_\psi^2$.

Let us now turn to the nonhydrodynamic modes \eqref{eq:mis-modes-0}.
For the fifth mode, up to first order in $\kappa_t$, we find
\begin{equation}\label{eq:sh-mode-5-1}
	\Omega_5 = -\frac{i}{\bulkr} + \alpha \shrc\left[ \frac{3\omega_\ell\omega_\perp}{\omega_\ell^2+\omega_\perp^2}\kappa_\psi
    +
    \left(1+
    \frac{3\omega_\ell^2}{\omega_\ell^2+\omega_\perp^2}
    \right)\kappa_\ell\right]\,.
\end{equation}
The term in brackets vanishes in the equilibrium-preserving $\psi$ direction, because there $\omega_\ell =\kappa_\ell = 0$.
Furthermore, the term of second order in momentum in this direction reads $\tfrac{4}{3}i\shc\kappa_\psi^2$.
This indicates that $\Omega_5$ is the counterpart of the nonhydrodynamic sound mode in the homogeneous case (with dispersion relation $\Omega=-i/\shr+4i\shc \kappa^2/3$), cf.\ Eq.\ \eqref{eq:hom-mis-disp2}.

For the other nonhydrodynamic modes, the terms of first order in momentum look more complicated, and we restrict our attention to their forms in the equilibrium-preserving $\psi$ direction.
In this direction, the first- and second-order terms in $\kappa_\psi$ of the sixth mode $\Omega_6$ vanish. 
This mode is the counterpart of the nonhydrodynamic shear mode in the homogeneous case (with dispersion relation $\Omega = -i / \shr$), cf.\ Eq.\ \eqref{eq:hom-mis-disp}.
The seventh mode $\Omega_7$, on the other hand, has nonvanishing terms of order $\kappa_\psi^2$, and reads
\begin{equation}
    \Omega^\eqp_{7} = \frac{-i+ \mcC_\perp}{\shr} + i \frac{(1+\mcC_\perp^2)\shc+6\alpha^2\shr\shrc^2}{(1+\mcC_\perp^2)^2}\kappa_\psi^2
    -\frac{\mcC_\perp^2(1+\mcC_\perp^2)\shc-3(1-\mcC_\perp^2)\alpha^2\shr\shrc^2}{\mcC_\perp(1+\mcC_\perp^2)^2}\kappa_\psi^2\,.
\end{equation}
Therefore, one can recognize this mode as the modification of the remaining nonhydrodynamic shear mode in the homogeneous case (with dispersion relation $\Omega=-i/\shr+i\shc\kappa^2$), cf.\ Eq.\ \eqref{eq:hom-mis-disp}.
However, this is not the only mode that has this homogeneous counterpart.
The eighth and ninth modes differ from the seventh only in the leading term in the equilibrium-preserving direction $\psi$,
\begin{equation}
	\Omega^{\eqp}_{8,\,9} =   \frac{-i\pm 2 \mcC_\perp}{\shr} + i \frac{(1+\mcC_\perp^2)\shc+6\alpha^2\shr\shrc^2}{(1+\mcC_\perp^2)^2}\kappa_\psi^2
    -\frac{\mcC_\perp^2(1+\mcC_\perp^2)\shc-3(1-\mcC_\perp^2)\alpha^2\shr\shrc^2}{\mcC_\perp(1+\mcC_\perp^2)^2}\kappa_\psi^2\,.
\end{equation}

Let us now consider the short-wavelength regime $\kappa_t \gg 1$.
In this limit, similar to the previous subsection, the symmetry of the modes is restored and we have
\begin{equation}
	\Re\Omega_{\rm nonhydro} \sim \pm \kappa_t \sqrt{\frac{\shc}{\shr}}\,,\qquad
	\Re\Omega_{\rm sound} \sim \pm \kappa_t\sqrt{\frac{4\shc + \shr}{3\shr}}\,.
\end{equation}
This means that the asymptotic group velocity does not exceed the speed of light if the standard linear causality condition, $\shr > 2\shc$ is satisfied. 
At this point, we turn to stability analysis of conformal MIS theory in inhomogeneous configurations.
To this end, we first consider the characteristic equation in the purely accelerating configuration \eqref{eq:acc-conf}. 
In this case, the characteristic equation decouples, as in homogeneous configurations, into two independent parts: the shear and sound channels. 
There is one nonpropagating mode, which is exactly equal to its homogeneous counterpart, i.e.,  $\Omega = - i / \shr$.
The remaining shear modes are modified by acceleration and found from the roots of
\begin{equation}\label{eq:acc-conformal-shear}
	\shr \Omega^2 + i\Omega - \shc\kappa_t^2 + 3 i \alpha\shrc\kappa_\ell = 0\,.
\end{equation}
In order for $\Im\Omega \leq 0$, $\shr$ must be positive and
\begin{equation}\label{eq:acc-shear-sc}
	\shc \kappa^2 + \left(\shc-9\shr\alpha^2\shrc^2\right)\kappa_\ell^2 > 0\,.
\end{equation}
If we restrict the momenta to the equilibrium-preserving directions by setting $\kappa_\ell=0$, this condition is satisfied if $\shc>0$.
On the other hand, for a mode with $\kappa=0$, the condition \eqref{eq:acc-shear-sc} requires 
\begin{equation}
	\alpha < \alpha_c \equiv \sqrt{\frac{\shc/\shr}{\abs{4\shc-\shr}}}\,,
\end{equation}
where we used Eq.\ \eqref{eq:shear-coeffs} for $\shrc$.
For any reasonable choice of transport coefficients, $\alpha_c$ is very large. 
For example, if we consider the parameters of Ref.\ \cite{Bhattacharyya:2007vjd},
\begin{equation}
	\shr = \frac{2-\ln 2}{2\pi}\,,\qquad 
	\shc = \frac{1}{4\pi}\,,
\end{equation}
we find
$
	\alpha_c \approx 5.61
$.
This value for $\alpha_c$ corresponds to a macroscopic length scale $a^{-1}$ that is much smaller than the typical microscopic length scale, which for uncharged conformal fluids is $T^{-1}$.
Consequently, the stability and causality conditions for homogeneous configurations, $\shr> 2 \shc>0$, guarantee the stability of the shear modes in the purely accelerating configuration, if the condition \eqref{eq:size-condition} is fulfilled.

Up to this point, every characteristic equation that we have considered reduces to its homogeneous counterpart in the equilibrium-preserving directions.
However, the characteristic equation of the sound channel in the purely accelerating configuration features a novel phenomenon: it is affected by acceleration even in the equilibrium-preserving directions.
This is because, in Eq.\ \eqref{eq:mis-shear-eom-sc}, $\alpha$ appears not only in the coefficients of $\d u_\ell$ but also of $\d u_\kappa$ and $\d u_\chi$.
With $\kappa_\ell = 0$, it reads
\begin{equation}\label{eq:acc-conf-sound}
    3\Omega^3\left(1-i\Omega\shr\right)^2 -
    \Omega\kappa^2 \left\{\left(1-i\Omega\shr\right)\left[1-i\left(7\shc+\shr\right)\Omega\right]-18\alpha^2\shrc^2\right\}
    -i \shc\kappa^4\left[1-i\left(4\shc+\shr\right)\Omega\right] = 0\,.
\end{equation}
The imaginary parts of some roots of this equation can be positive if $\alpha$ is larger than a critical value.
As we have already restricted the momenta to the equilibrium-preserving directions, it might be tempting to conclude that conformal MIS theory could become unstable in the purely accelerating configuration.
However, further inspection shows that such a critical value of $\alpha$ is always larger than one, violating the condition \eqref{eq:size-condition}.
In order for the imaginary parts of the roots of the characteristic equation of the sound channel in the equilibrium-nonpreserving $\ell$ direction to be positive, similarly large values of $\alpha$ are required.

We close this section by commenting on the stability of conformal MIS theory in the rigidly rotating configuration \eqref{eq:rigid-rotation}.
In this case, the characteristic equation remains of order nine and is thus quite complicated even after restricting the momenta to the equilibrium-preserving  $\psi$ direction.
We insert $a = v_\varphi \omega_\perp$, where $v_\varphi = \rho\Omega_0$, into the characteristic equation and perform a Routh-Hurwitz analysis.
Consequently, we find that $\Im\Omega$ can be positive even with momenta restricted to the equilibrium-preserving  $\psi$ direction, if $\Omega_0 \gg T_0$, or $v_\varphi$ is very close to the speed of light.
The former case violates the condition \eqref{eq:size-condition}, while the latter one corresponds to radii very close to the causal boundary of the fluid.
Therefore, we conclude that in the domain of validity of MIS hydrodynamics, the stability conditions found for homogeneous configurations extend to accelerating and rigidly rotating configurations.

\section{Concluding remarks}
\label{sec:conclusion}
We have proposed a method to find local plane-wave solutions to the linearized hydrodynamic equations of motion in inhomogeneous equilibrium configurations, i.e., configurations with nonzero thermal vorticity. 
Our method is based on extending the perturbations of the conserved currents around the equilibrium configuration to the tangent bundle using Wigner transforms, and then Fourier transforming them to the cotangent bundle. 
\amm{The tangent bundle plays the role of a homogeneous equilibrium configuration where, in an infinitesimal neighborhood, the equilibrium quantities are constant.
By Fourier transforming, we choose the solutions to the equations of motion that are superpositions of linear waves in this infinitesimal domain.}
This procedure leads to a homogeneous system of linear equations, from which, by setting its determinant to zero, one finds the linear modes in the inhomogeneous configurations. 

Contrary to homogeneous equilibrium configurations, a positive sign of the imaginary parts of the modes in the inhomogeneous case does not necessarily indicate a linear instability. 
This is because the frequencies of the modes depend on the local quantities in equilibrium. 
In flat space-time, the latter do not change in the directions perpendicular to the thermal vorticity. 
We refer to these directions as equilibrium-preserving directions.
We showed that these directions exist, if space-time is flat and the kinematic vorticity is perpendicular to the acceleration.
Restricting the momenta of the modes to these equilibrium-preserving directions, if the imaginary part of at least one mode is positive, an instability exists.
Such an instability is, however, only physically relevant as long as the length scale related to the thermal vorticity remains much larger than the typical microscopic scale of the system.
On the other hand, a positive imaginary part of a mode with nonvanishing momenta in an equilibrium-nonpreserving direction does not necessarily prove the instability of the system.

As an application, we considered MIS hydrodynamics.
We first studied a fluid for which the bulk viscous pressure is the only source of dissipation. 
We showed that coupling between the bulk viscous pressure and the acceleration leads to novel contributions to the dispersion relations of the sound modes in the direction of acceleration.
Consequently, the group velocities of the sound modes in this direction are asymmetrically modified in the long-wavelength regime.
However, in the short-wavelength regime, symmetry is recovered and the group velocities remain smaller than the speed of light if the theory is linearly causal. 
In the equilibrium-preserving directions, the novel contributions vanish, and the standard stability conditions of MIS theory for the case of bulk viscosity only are recovered.
On the other hand, in the direction of acceleration, the imaginary part of one of the modes can become positive if the magnitude of the acceleration is sufficiently large.
However, we have argued that the corresponding large accelerations can neither be physically realized nor are in the domain of validity of MIS hydrodynamics.

Finally, we have considered a conformal fluid in MIS theory.
In this case, not only is the dispersion relation of the modes modified by the thermal vorticity, but also the number of modes is increased to nine in the presence of rotation.
In the short-wavelength regime, the asymmetry of the modes is eliminated and the standard condition for linear causality is recovered. 
In contrast to the case of bulk viscosity only, these modes have novel contributions even when the momenta are restricted to the equilibrium-preserving directions. 
Consequently, the imaginary parts of at least one mode can be positive for sufficiently strong thermal vorticities.
However, such an effect, with a reasonable choice of parameters, only occurs beyond the validity of the hydrodynamic theory. 
This is either when the microscopic and macroscopic scales are similar or when boundary effects cannot be neglected.
Consequently, we conclude that MIS theory in its domain of validity remains linearly stable in inhomogeneous configurations, with the standard stability and causality conditions.
This conclusion agrees with Ref.\ \cite{Hiscock:1983zz}, which uses the so-called information current method. 
We note that, although this method does not assume a homogeneous equilibrium configuration, it neglects the existence of boundaries, which are always present in inhomogeneous equilibrium configurations.

The methods introduced here can be applied to different hydrodynamic theories to find linear waves in inhomogeneous equilibrium configurations. 
Hydrodynamic theories with quantum corrections arising from acceleration and rotation \cite{Palermo:2021hlf,Becattini:2020qol,Ambrus:2021eod,Ambrus:2019cvr,Prokhorov:2019yft} and formulations of spin hydrodynamics that explicitly contain the thermal vorticity \cite{Weickgenannt:2022zxs} are of particular interest.
This work can also be extended by an investigation of boundary and size effects on mode propagation and stability in inhomogeneous equilibrium configurations. 
\amm{The Wigner-Fourier transforms that were utilized in this work might also be used to study hydrodynamic fluctuations as an alternative to the method introduced in Ref.\ \cite{An:2019osr}.}

\begin{acknowledgments}
	M.\ S.\ thanks L.\ Gavassino, V.\ Ambrus, A.\ Palermo, D.\ Wagner, and A.\ Dash for fruitful discussions.
	This work is supported by the Deutsche Forschungsgemeinschaft
	(DFG, German Research Foundation) through the Collaborative Research
	Center CRC-TR 211 ``Strong-interaction matter under extreme conditions''
	- project number 315477589 - TRR 211 and by the State of Hesse within the Research Cluster
	ELEMENTS (Project ID 500/10.006).
\end{acknowledgments}

\appendix
\section{\label{app:Example}Rigidly rotating fluid in a Schwarzschild metric}
%%%%%%%%
In this appendix, we consider a rigidly rotating fluid in Schwarzschild metric, whose line element in spherical coordinates $(t,r,\theta,\phi)$ reads
\begin{eqnarray}
	\dd{s}^2 = \left(1-\frac{r_s}{r}\right)\dd{t}^2-\left(1-\frac{r_s}{r}\right)^{-1}\dd{r}^2 - r^2\dd{\Omega_s^2}\,,
	\label{eq:sch_ds2}
\end{eqnarray}
where $r_s=2GMr$ is the Schwarzschild radius and $\dd{\Omega_s}=\dd{\theta}^2+\sin^2\theta\dd{\phi}^2$. 
This configuration is found by assuming \cite{Ambrus:2016ocv}
\begin{equation}
	\beta =  \inv{T_0}\left(\pdv{t}+\Omega_0\pdv{\phi}\right)\,,
\end{equation}
where $\Omega_0$ is a constant of dimension energy. 
The above $\beta$-vector is timelike if
\[
	1 - \frac{r_s}{r} - \Omega_0^2 r^2 \sin^2 \theta   > 0\,.
\]
In spherical coordinates, the velocity and temperature are given by
\begin{equation}
	u^\mu = \gamma\left(1,0,0, \Omega_0\right)\,,
	\qquad
	T = \gamma T_0\,,
\qq{with}
	\gamma = \frac{1}{\sqrt{1 - r_s/r - \Omega_0^2 r^2 \sin^2 \theta  }}\,.
\end{equation}
As in the rotating equilibrium configuration \eqref{eq:rigid-rotation}, both acceleration and kinematic vorticity are nonzero,
\begin{subequations}
\begin{eqnarray}
    a^\mu &=& -\half\gamma^2 \left(0, \frac{(r-r_s)(2\Omega_0^2r^3\sin^2\theta-r_s)}{r^3}, \Omega_0^2\sin 2\theta,0\right)\,,
    \\
    \omega^\mu &=& \gamma^2 \left(0, \frac{(r-r_s)\Omega_0\cos\theta}{r},-\frac{(2r-3r_s)\Omega_0\sin\theta}{2r^2},0\right)\,.
\end{eqnarray}
\end{subequations}
These vectors are not orthogonal, 
\begin{equation}
    \omega\cdot a = - \gamma^2\frac{r_s\Omega_0\cos\theta}{2r^2}\,.
\end{equation}
We note that even with a vanishing $\Omega_0$, the equilibrium configuration is inhomogeneous due to gravity, as required by the Tolman law, i.e, $T=T_0/\sqrt{g_{00}}$. 
In this limit, 
\begin{equation}
    u^\mu = \gamma (1,\vb{0})\,,
    \qquad
    T = \gamma T_0\,,
    \qquad
    a^\mu = \left(0, \frac{r_s}{2r^2}, 0,0\right)\,,
\end{equation}
with $\gamma = 1/\sqrt{1-r_s/r} = 1/\sqrt{g_{00}}$. 
As expected, the kinematic-vorticity vector vanishes in this case.

\section{\label{app:h-lift}Proof of identities for the tangent bundle}

In this appendix, we prove some identities regarding the horizontal lift in the tangent bundle, including Eq.\ \eqref{eq:yd-to-hl}.
First, we realize that $\hlift_\mu y^\nu=0$ since
\begin{eqnarray}\label{eq:hlift-y}
    \hlift_\mu y^\nu &=& \left(\nabla_\mu  - \christoffel{\sigma}{\mu}{\rho}y^\rho\p^y_\sigma\right)y^\nu =
    \christoffel{\nu}{\mu}{\rho}y^\rho - \christoffel{\sigma}{\mu}{\rho}y^\rho\delta^\nu_\sigma 
    = 0\,.
\end{eqnarray}
Furthermore, $\p^y$ commutes with $\hlift$, 
\begin{eqnarray}\label{eq:hlift-comm-dy}
	\comm{\hlift_\mu}{\p^y_{\nu}} &=& \comm{\nabla_\mu-\christoffel{\a}{\mu}{\b}(x)y^\b\p^y_{\a}}{\p^y_{\nu}}= \comm{\nabla_\mu}{\p^y_{\nu}}-\christoffel{\a}{\mu}{\b}\comm{y^\b\p^y_{\a}}{\p^y_{\nu}}
	= -\christoffel{\a}{\mu}{\nu}\p^y_{\a}
    + \christoffel{\a}{\mu}{\b}\delta^\b_\nu \p^y_{\a}
	=0\,.
\end{eqnarray}
Note that, similarly to Eq.\ \eqref{eq:hlift-y},
\begin{eqnarray}\label{eq:chlift-k}
    \chlift_\mu k_\nu &=& \left(\nabla_\mu  + \christoffel{\sigma}{\mu}{\rho}k_\sigma\p_k^\rho\right)k_\nu=
    -\christoffel{\rho}{\mu}{\nu}k_\rho + \christoffel{\sigma}{\mu}{\rho}k_\sigma \delta^\rho_\nu 
    = 0\,.
\end{eqnarray}
Also, we have
\begin{eqnarray}\label{eq:chlift-comm-dk}
	\comm{\chlift_\mu}{\p_k^{\nu}} &=& \comm{\nabla_\mu+\christoffel{\a}{\mu}{\b}(x)k_\a\p_k^{\b}}{\p_k^{\nu}}=\left[\nabla_\mu,\p_k^{\nu}\right]+\christoffel{\a}{\mu}{\b}\comm{k_\a\p_k^{\b}}{\p_k^{\nu}}
	= \christoffel{\nu}{\mu}{\rho}\p_k^{\rho}-\christoffel{\a}{\mu}{\b}\delta^\nu_\a\p_k^\b
	=0\,.
\end{eqnarray}
Using Eq.\ \eqref{eq:hlift-comm-dy}, we find that commuting $y \cdot\hlift$ with $\p^y_{\mu}$ generates a horizontal lift,
\begin{eqnarray}\label{eq:ydothlift-comm-dy}
	\comm{y \cdot \hlift}{\p^y_{\mu}} &=& -\comm{\p^y_{\mu}}{y^\nu}\hlift_\nu - y^\nu\comm{\p^y_{\mu}}{\hlift_\nu}
	= -\hlift_\mu\,.
\end{eqnarray}
We define the full curvature in the tangent bundle as the commutator of two horizontal lifts \cite{Liu:2018xip}, 
\begin{equation}
	G_{\mu\nu}\, \mcK^{\a_1\cdots}_{\b_1\cdots}(x,y) \equiv \left[\hlift_\mu,\hlift_\nu\right]\mcK^{\a_1\cdots}_{\b_1\cdots}(x,y)\,,
\end{equation}
where $\mcK^{\a_1\cdots}_{\b_1\cdots}$ is a tensor of arbitrary rank. 
To identify the $y$-dependent part of the total curvature, we consider its action on a scalar function $F(x,y)$,
\begin{eqnarray}\label{eq:scalar-full-curv}
	\left[\hlift_\mu,\hlift_\nu\right] F(x,y)\ &=&  \Big[\nabla_\mu-\christoffel{\beta}{\mu}{\alpha}y^\alpha\p^y_\beta,\,
	{\nabla_\nu}-\christoffel{\sigma}{\nu}{\rho}y^\rho\p^y_\sigma\Big]F(x,y)\,
	%%%%%%%%%%%%%%%%%%%%%%%%%%%%%%%%%%%%%%%%%%%%%%%%%%%%%%%
	\nn\\
    &=& - \Big[\partial_\mu,\christoffel{\sigma}{\nu}{\rho}y^\rho\p^y_\sigma\Big]F(x,y)
	% \nn\\
	% &&\hspace{-3cm}
	- \Big[\christoffel{\b}{\mu}{\a}y^\a\p^y_\b,\partial_\nu\Big]F(x,y)
	+ \Big[\christoffel{\b}{\mu}{\a}y^\a\p^y_\b,\christoffel{\sigma}{\nu}{\rho}y^\rho\p^y_\sigma\Big]F(x,y)\,
	%%%%%%%%%%%%%%%%%%%%%%%%%%%%%%%%%%%%%%%%%%%%%%%%%%%%%%%
	\nn\\
    &=& - R^\sigma_{\r\mu\nu}y^\r\p^y_\sigma F(x, y)\,,
\end{eqnarray}
where we have used 
\begin{equation}
    R^\sigma_{\r\mu\nu} = 2\left(\partial_{[\mu}\christoffel{\sigma}{\nu]}{\rho}
+\christoffel{\sigma}{\beta}{[\mu}\christoffel{\beta}{\nu]}{\r}\right)\,.
\end{equation}
For tensors of arbitrary rank, commuting $\left[\nabla_\mu,\nabla_\nu\right]$ gives rise to additional curvature terms. 
However, the $y$-dependent part is independent of the tensor rank as is the same as in Eq.\ \eqref{eq:scalar-full-curv}.

Using Eq.\ \eqref{eq:scalar-full-curv}, we prove Eq.\ \eqref{eq:yd-to-hl}.
In order to do so, we start with \cite{fonarev}
\begin{equation}\label{eq:bch-identity}
	\left[\hat{A},e^{\hat{B}}\right] = - \sum_{n=1}^{\infty}\inv{n!}\overbrace{\left[\hat{B},\left[\hat{B}, \left[\cdots\left[\hat{B},\hat{A}\right]\cdots\right]\right]\right]}^{\text{n times}}\, e^{\hat{B}}\,.
\end{equation}
This identity can be written using the so-called adjoint map $\mcC$ in a compact form. 
The adjoint map is defined as
\begin{equation}
    \adm{\hat{X}}\hat{Y} \equiv \left[X,Y\right]\,,
\end{equation}
were $\hat{X}$ and $\hat{Y}$ are some operators.
Consequently, Eq.\ \eqref{eq:bch-identity} is rewritten as
\begin{equation}\label{eq:bch-identity-adm}
	e^{\hat{B}}\hat{A} = \Big\{  e^{\adm{\hat{B}}}\hat{A} \Big\}  e^{\hat{B}}\,.
\end{equation}
Now, let us consider acting both sides of this identity on a scalar $F(x)$ for the following operators
\[
	\hat{A}\to\p^y_\mu\,,\qquad\hat{B} \to y\cdot\hlift\,,
\]
which gives rise to
\[
	e^{\adm{y\cdot\hlift}}\p^y_\mu e^{y\cdot\hlift} F(x) = e^{y\cdot\hlift} \p^y_\mu F(x) = 0\,.
\]
Therefore, with Eq.\ \eqref{eq:ydothlift-comm-dy},
\begin{eqnarray*}
	0 &=& e^{\adm{y\cdot\hlift}}\p^y_\mu F(x,y)
	\nn\\
	&=& \left[1+\adm{y\cdot\hlift}+\sum_{n=2}^{\infty}\frac{\adm{y\cdot\hlift}^n}{n!}\right]\p^y_\mu F(x,y)
	\nn\\
	&=&\p^y_\mu F(x,y)-\hlift_\mu F(x,y) +\sum_{n=2}^{\infty}\frac{\adm{y\cdot\hlift}^{n-2}}{n!}\adm{y\cdot\hlift}\adm{y\cdot\hlift}\p^y_\mu F(x,y)
	\nn\\
	&=&\p^y_\mu F(x,y)-\hlift_\mu F(x,y) -\sum_{n=2}^{\infty}\frac{\adm{y\cdot\hlift}^{n-2}}{n!}\adm{y\cdot\hlift}\hlift_\mu F(x,y)
	\nn\\
	&=& \p^y_\mu F(x,y)-\hlift_\mu F(x,y) -y^\nu\sum_{n=0}^{\infty}\frac{\adm{y\cdot\hlift}^{n}}{(n+2)!}\left[\hlift_\nu,\hlift_\mu\right] F(x,y)
	\nn\\
	&=&\p^y_\mu F(x,y)-\hlift_\mu F(x,y) + y^\nu\sum_{n=0}^{\infty}\frac{\adm{y\cdot\hlift}^{n}}{(n+2)!}G_{\mu\nu} F(x,y)\,,
\end{eqnarray*}
which yields Eq. \eqref{eq:yd-to-hl}.

\section{\label{app:hlift-to-chlift}
Direct proof of Eq.\ \eqref{eq:hlift-to-chlift}}

The identity \eqref{eq:hlift-to-chlift} can be directly proved as follows. 
Since the $y$- and $k$-dependent part of the horizontal lifts are independent of the index structure of a tensor, we prove the identity for a scalar $F(x,y)$,
\begin{eqnarray}
    \hlift_\mu F(x,y) &=& \hlift_\mu \int_k e^{-ik\cdot y} F(x,k)
    \nn\\&=&
     \int_k \left(\nabla_\mu -\christoffel{\rho}{\mu}{\nu} y^\nu\p^y_\rho\right)\left[ e^{-ik\cdot y} F(x,k)\right]
    \nn\\&=&
     \int_k \left[\nabla_\mu F(x,k) +iF(x,k)\christoffel{\rho}{\mu}{\nu} y^\nu k_\rho\right] e^{-ik\cdot y}
     \nn\\&=&
     \int_k \left[\nabla_\mu F(x,k) -F(x,k)\christoffel{\rho}{\mu}{\nu}  k_\rho\p_k^\nu\right] e^{-ik\cdot y}\,,
\end{eqnarray}
where in the second line we have used the invariance of the volume element. 
We have also used the fact that $k\cdot y$ is a scalar, and therefore $\nabla_\mu \left(k\cdot y\right) = \p_\mu \left(k\cdot y\right) = 0$. 
Now, by performing an integration by parts, we find
\begin{eqnarray}
    \hlift_\mu F(x,y) &=& \int_k e^{-ik\cdot y}\left[\nabla_\mu F(x,k) +\christoffel{\rho}{\mu}{\nu} k_\rho\p_k^\nu F(x,k)\right]
    \nn\\
    &=& \int_k e^{-ik\cdot y}\chlift_\mu F(x,k) 
    \,,
\end{eqnarray}
which completes the proof. 
One can convince oneself that this proof is independent of the rank of the tensor $F(x)$. 
\section{\label{app:voriticty}Covariant derivative of thermal vorticity}

To find $\nabla_\mu \varpi_{\alpha\beta}$, we start by writing
\begin{equation}
	\nabla_\mu \varpi_{\alpha\beta} = -\nabla_\mu \nabla_\alpha \beta_\beta, \qquad 
	\nabla_\alpha \varpi_{\mu \beta} = -\nabla_\alpha \nabla_\mu \beta_\beta\,,
\end{equation}
where we have used the Killing condition \eqref{eq:killing} to rewrite the thermal vorticity as a single covariant derivative. 
Then, we subtract these two equations, and use the definition of the Riemann tensor, to find
\begin{equation}
	\nabla_\mu \varpi_{\alpha\beta} - \nabla_\alpha \varpi_{\mu \beta} = [\nabla_\alpha, \nabla_\mu] \beta_\beta \equiv R_{\beta\sigma \alpha\mu} \beta^\sigma =  R^\sigma{}_{\beta\mu\alpha} \beta_\sigma\,.
\end{equation}

Permuting the indices clockwise we obtain, 
	\begin{equation}
		\nabla_\alpha\varpi_{\beta\mu} - \nabla_\beta \varpi_{\alpha\mu} =  R^\sigma{}_{\mu\alpha\beta} \beta_\sigma, \qquad 
				\nabla_\beta \varpi_{\mu\alpha} - \nabla_\mu \varpi_{\beta\alpha} =  R^\sigma{}_{\alpha\beta\mu} \beta_\sigma\,,
			\end{equation}
Adding the three equations gives rise to
	\begin{equation}\label{eq:nabla-varpi}
				\nabla_\mu \varpi_{\alpha\beta} =  \frac{1}{2}(R^\sigma{}_{\beta\mu\alpha} - R^\sigma{}_{\mu\alpha\beta} + R^\sigma{}_{\alpha\beta\mu}) \beta_\sigma = -R^\sigma{}_{\mu\alpha\beta} \beta_\sigma = R_{\alpha\beta\mu\sigma} \beta^\sigma\,,
\end{equation}
where we have used the cyclic property $R^\sigma{}_{\alpha\beta\mu} + R^\sigma{}_{\mu\alpha\beta} + R^\sigma{}_{\beta\mu\alpha} = 0$, the symmetry relation $R_{\sigma\alpha \beta\mu} = R_{\beta\mu \sigma\alpha}$, and the antisymmetry relations $R_{\sigma\alpha \beta\mu} = -R_{\alpha\sigma \beta\mu} = -R_{\sigma\alpha \mu\beta}$.

\bibliography{main}
\end{document}